\documentclass[10pt]{article}
\usepackage{bbm}
\usepackage[final]{graphics}
\usepackage{amsmath}
\usepackage{amsfonts,amsbsy}
\usepackage{amssymb}
\usepackage{bm}
\usepackage{color}

\def\empile#1\over#2{\mathrel{\mathop{\kern 0pt#1}\limits_{#2}}}
\def\bs{\boldsymbol}

\def\TODO#1{}

\newcommand{\slL}{\raise.15ex\hbox{$/$}\kern-.53em\hbox{$L$}}
\newcommand{\slP}{\raise.15ex\hbox{$/$}\kern-.53em\hbox{$P$}}
\newcommand{\slD}{\raise.15ex\hbox{$/$}\kern-.67em\hbox{$D$}}
\newcommand{\slp}{\raise.1ex\hbox{$/$}\kern-.63em\hbox{$p$}}
\newcommand{\slq}{\raise.1ex\hbox{$/$}\kern-.53em\hbox{$q$}}
\newcommand{\slv}{\raise.1ex\hbox{$/$}\kern-.63em\hbox{$v$}}
\newcommand{\slR}{\raise.15ex\hbox{$/$}\kern-.53em\hbox{$R$}}
\newcommand{\slQ}{\raise.15ex\hbox{$/$}\kern-.53em\hbox{$Q$}}
\newcommand{\slK}{\raise.15ex\hbox{$/$}\kern-.53em\hbox{$K$}}
\newcommand{\slk}{\raise.15ex\hbox{$/$}\kern-.53em\hbox{$k$}}
\newcommand{\slSigma}{\raise.15ex\hbox{$/$}\kern-.53em\hbox{$\Sigma$}}
\newcommand{\slcalP}{\raise.15ex\hbox{$/$}\kern-.63em\hbox{$\cal P$}}
\newcommand{\slcalA}{\raise.15ex\hbox{$/$}\kern-.63em\hbox{$\cal A$}}
\newcommand{\slA}{\raise.15ex\hbox{$/$}\kern-.73em\hbox{$A$}}
\newcommand{\slbfA}{\raise.15ex\hbox{$/$}\kern-.73em\hbox{${\imb A}$}}
\newcommand{\slpartial}{\raise.15ex\hbox{$/$}\kern-.53em\hbox{$\partial$}}
\newcommand{\sla}{\raise.15ex\hbox{$/$}\kern-.53em\hbox{$a$}}
\newcommand{\slb}{\raise.15ex\hbox{$/$}\kern-.53em\hbox{$b$}}
\newcommand{\slc}{\raise.15ex\hbox{$/$}\kern-.53em\hbox{$c$}}
\newcommand{\slC}{\raise.15ex\hbox{$/$}\kern-.63em\hbox{$C$}}
\newcommand{\sln}{\raise.15ex\hbox{$/$}\kern-.575em\hbox{$n$}}

\newcommand\ontop[2]{\genfrac{}{}{0pt}{}{#1}{#2}}

\begin{document}

\title{\bf Lattice worldline representation\\ of correlators in a background field}
\author{Thomas Epelbaum${}^1$, Fran\c cois Gelis${}^2$, Bin Wu${}^2$}
\maketitle

\begin{itemize}
\item[{\bf 1}.] McGill University, Department of Physics\\ 3600 University Street, Montreal QC H3A 2T8, Canada
\item[{\bf 2.}] Institut de Physique Th\'eorique, CEA/DSM/Saclay\\
91191 Gif sur Yvette, France
\end{itemize}

\begin{abstract}
  We use a discrete worldline representation in order to study the
  continuum limit of the one-loop expectation value of dimension two
  and four local operators in a background field. We illustrate this
  technique in the case of a scalar field coupled to a non-Abelian
  background gauge field. The first two coefficients of the expansion
  in powers of the lattice spacing can be expressed as sums over
  random walks on a $d$-dimensional cubic lattice. Using combinatorial
  identities for the distribution of the areas of closed random walks
  on a lattice, these coefficients can be turned into simple
  integrals. Our results are valid for an anisotropic lattice, with
  arbitrary lattice spacings in each direction.
\end{abstract}

\section{Introduction}
\subsection{Motivation}
The classical statistical approximation (CSA) is an approximate scheme
to study in real time the dynamics of a system of fields, as an
initial value problem. It has been used in cosmology
\cite{Son1,KhlebT1,TranbW1}, in high energy nuclear physics to study
heavy ion collisions \cite{BergeBSV1,BergeBSV2,EpelbG3}, in studies of
the Schwinger mechanism \cite{FukusGL1,GelisT2}, and in cold atom
physics \cite{Norri1,NorriBG1}.  In this approximation, the time
evolution of the fields is classical (i.e. deterministic), and one
averages over fluctuations of their initial conditions. Obviously,
this scheme neglects all the quantum effects that would normally
affect the time evolution of a system. In the path integral language,
it corresponds to taking the saddle point of the integral. This can be
justified if the typical field amplitude in the system under
consideration is large, so that the commutator of a pair of fields is
much smaller than the typical field squared.

Some quantum effects can nevertheless be included in the classical
statistical approximation via the fluctuations of the initial
fields. Indeed, it can be shown on general grounds that the leading
(i.e. of order $\hbar$) quantum effects come entirely from the initial
condition for the density operator of the system, while the quantum
corrections that alter its time evolution only start at the order
$\hbar^2$. In fact, there is a unique statistical ensemble of initial
classical fields such that the CSA coincides with the exact ${\cal
  O}(\hbar)$ result for all
observables~\cite{Son1,Gelis15,EpelbG2}. These initial fields have a
flat spectrum in momentum space, that extends to arbitrarily large
momenta.

Because the CSA implemented in this manner\footnote{Another common
  implementation of the CSA is to use a spectrum of field fluctuations
  with a compact (or at least falling faster than $1/k$) momentum
  spectrum \cite{BergeBSV1,BergeBSV2}.  This type of initial
  distribution corresponds to a classical ensemble of quasiparticle
  excitations, instead of quantum fluctuations. This version of the
  CSA is free of any ultraviolet divergences (for a spectrum that
  falls like $k^{-1}$, one gets some ultraviolet divergences, but the
  resulting approximation is super-renormalizable
  \cite{AartsS1,AartsS3}), but it also does not contain anything
  quantum. It may coincide with the underlying theory at tree level,
  but not beyond.}  contains all the ${\cal O}(\hbar)$ contributions
of the underlying theory, it also contains all their singularities,
and in particular the ultraviolet divergences. In addition, it
contains some of the higher order contributions, but not all of them
since the quantum corrections to the time evolution are missing in
this approximation. It has been shown recently that this leads to a
dependence on the ultraviolet cutoff that cannot be disposed of by the
usual renormalization of the parameters of the theory. This can be
seen in computer simulations using the CSA \cite{BergeBSV3}, by a
perturbative analysis of the graphs that arise in the CSA
\cite{EpelbGW1}, and also from the study of cutoff effects in the
classical approximation of the Boltzmann equation \cite{EpelbGTW1}.

At the moment, it is not known whether the CSA can be modified in
order to remove these unwanted terms. However, regardless of this
interesting theoretical question, it is important to have a good
understanding of the structure of the standard 1-loop ultraviolet
divergences. Indeed, since they are identical in the CSA and in the
exact theory, they can be removed by the usual renormalization
procedure. But their form may be quite complicated in the lattice
formulation of the CSA, especially for a generic lattice that may have
anisotropic lattice spacings\footnote{In applications to heavy ion
  collisions, it is common to have a much smaller lattice spacing in
  the direction of the collision axis.}. Generically, this requires
the calculation at one-loop of the expectation value of interest, with
lattice regularization and propagators, in the presence of a
non-Abelian background field. Unfortunately, lattice perturbation
theory is quite complicated, even for this seemingly simple task (see
\cite{Capit1} for a review). The main issue is the treatment of the
background field, and the fact that one recovers gauge invariant
results by combining several pieces that are not individually gauge
invariant (see the appendix \ref{app:LPT} for an example of such
calculation in the present context).

In the present paper, we pursue a different approach in order to
obtain these 1-loop quantities, based on the so-called worldline
formalism. Historically, this formalism emerged from ideas based on
the limit of infinite tension in string theory\footnote{An earlier
  example of this approach is a string-inspired calculation of the
  1-loop $\beta$ function of Yang-Mills theory \cite{MetsaT1}.}
\cite{BernK1,BernK2}, and it soon appeared that it provides a powerful
way of organizing field theory calculations, especially when gauge
symmetry is part of the game.  A pure field theory understanding of
this formalism was later proposed in ref.~\cite{Stras1}, by a method
which is similar to Schwinger's proper time representation. For
reviews on this approach, the reader may consult
refs.~\cite{Schub1,Schub2}. This formalism has been applied to the
evaluation of effective actions
\cite{ReuteSS1,SchmiS1,SchmiS2,GiesSV1}, to the study of pair
production by an external field \cite{DunneS1,DunneWGS1} or the
Casimir effect \cite{GiesLM1,GiesK1}. Numerical algorithms based on
this formalism have also been proposed
\cite{GiesL1,GiesL2,GiesK1}. Since our goal is to apply this formalism
to a lattice field theory, the closest work we are aware of is in
refs.~\cite{SchmiS3,SchmiS4}, where a new method for evaluating
functional determinants in terms of worldlines was proposed.

In the present paper, we use this formalism in order to obtain useful
expressions for 1-loop expectations values in a lattice field theory
coupled to a (fixed) gauge background.  As we shall see, the worldline
formalism is well suited for this application because it enables one
to have only gauge invariant objects at all stages of the calculation.
Then, we use these expressions in order to study the limit of small
lattice spacing. In this limit, we obtain an expansion in terms of the
background field strength, with coefficients that are given by sums
over closed loops on the lattice, weighted by powers of the area
enclosed by the loop. Thus, the worldline formalism relates the
coefficient of the short distance expansion to combinatorial
properties of closed loops on a cubic lattice.

In the simple case where the lattice spacings are the same in all the
directions, the combinatorial formulas we need were already known and
can be found in ref.~\cite{MingoN1}. The generalization to anisotropic
lattice spacings requires some combinatorial formulas that we could
not find in the literature. A numerical exploration of all the random
walks of length $\le 20$ led us to conjecture a number of such
formulas (discussed in the appendix \ref{app:area-aniso}), that extend
and generalize those of ref.~\cite{MingoN1}.  A proof of these
formulas\footnote{Eqs.~(\ref{eq:area-nxy-00}) and
  (\ref{eq:area4-nxy-00}) can be viewed as ``unsummed'' versions of
  the formulas (1.5) and (1.6) of ref.~\cite{MingoN1}, while
  eqs.~(\ref{eq:area-nxy-0x}), (\ref{eq:area-nxy-02x}),
  (\ref{eq:area-nxy-0xy}), (\ref{eq:area6-nxy-00}),
  (\ref{eq:area8-nxy-00}) and (\ref{eq:area10-nxy-00}) seem to be
  totally new. The appendix \ref{app:LPT} provides an indirect proof
  of eq.~(\ref{eq:area-nxy-00}), since we rederive the expansion of
  $\big<\phi_a^*(0)\phi_a(0)\big>$ --that relies on
  eq.~(\ref{eq:area-nxy-00}) in the worldline approach-- using lattice
  perturbation theory. } is presented in a separate paper,
ref.~\cite{EpelbGW3}.

\subsection{Model}
In order to keep things rather simple and focus on the main aspects of
the worldline formalism, we consider a complex scalar field coupled to
an external non-Abelian gauge field. This background field is given
once for all, and does not fluctuate. We neglect the self-interactions
of the scalar field. The Lagrangian reads~:
\begin{equation}
{\cal L}\equiv \sum_{\mu=1}^d (D_\mu \phi)^*(D^\mu\phi)\; ,
\label{eq:L}
\end{equation}
where $D_\mu\equiv\partial_\mu-igA_\mu$ is the covariant derivative in
the presence of the background field. We also assume that the system
is initialized at $x^0=-\infty$ into the perturbative vacuum.

We consider the expectation values of local gauge invariant operators
made of the field $\phi$, e.g. $\phi^*\phi$, $\phi^*D_\mu D^\mu\phi$,
$(D_\mu\phi)^*(D_\nu\phi)$. With the Lagrangian given in
eq.~(\ref{eq:L}) and the vacuum state as initial condition, these
expectation values are given by a 1-loop graph in a background
field. These loops contain a pure vacuum contribution which is
ultraviolet divergent. In addition, for operators that have a
sufficiently high dimension, there may be subleading ultraviolet
divergences, whose structure is however strongly constrained by gauge
invariance. Our goal in this paper is to investigate the structure of
these divergences, with a lattice regularization.

We work with an Euclidean metric, in $d$ space-time dimensions, on a
discrete cubic lattice. For definiteness, the spatial directions are
chosen to be $1,\cdots ,d-1$ and the direction $d$ can be considered as
the time direction (although this distinction is hardly relevant with
an Euclidean metric). Our goal is to study the general case of
arbitrary lattice spacings $a_1,\cdots,a_d$ in each direction, but in
the sections \ref{sec:phi2} and \ref{sec:bilocal} we expose the
formalism with an isotropic lattice for simplicity. The coordinates
are labeled $x_1$ to $x_d$, and we denote by $\hat{1},\cdots,\hat{d}$
the vectors corresponding to one lattice spacing in each of the
directions. Therefore, any point on the lattice can be represented as
$x=x_1\hat{1}+\cdots+x_d \hat{d}$, where the $x_i$ are integers.

\subsection{Preview of the results}
In $d$ dimensions and for completely arbitrary lattice spacings in
each direction, the expectation value of $|\phi(0)|^2$ at 1-loop in a
background field admits the following expansion in powers of the
lattice spacings,
\begin{eqnarray}
\big<\phi_a^*(0)\phi_a(0)\big>
&\empile{=}\over{\{a_i\to 0\}}&
\frac{{\bf a}^2}{2d \prod_{i=1}^d a_i}\Bigg[
{\rm tr}_{\rm adj}(1)\,{\bs C}_0
\nonumber\\
&&\qquad
-
\frac{g^2}{4}\;\sum_{i<j} a_i^2 a_j^2 F^{ij}_a(0)F^{ij}_a(0)
\;{\bs C}_4^{ij;ij}+\cdots
\Bigg]\; ,
\end{eqnarray}
with
\begin{equation}
{\bs C}_0\equiv \int_0^\infty dt\;e^{-t}\;\prod_{r=1}^d I_0(\tfrac{h_r t}{d})
\end{equation}
and
\begin{equation}
{\bs C}_4^{ij;ij}=\frac{h_i h_j}{12d^2}
\int_0^\infty dt\; e^{-t}\;t^2\;\Big[\prod_{k\not=i,j}I_0(\tfrac{h_kt}{d})\Big]
\;I_1(\tfrac{h_it}{d})\;I_1(\tfrac{h_jt}{d})\; .
\end{equation}
$I_0$ and $I_1$ are modified Bessel functions of the first kind.  In
all these equations, we denote
\begin{equation}
{{\bf a}^{-2}}\equiv \frac{1}{d}\sum_{i=1}^d a_i^{-2}\qquad  h_i\equiv \frac{{\bf a}^2}{a_i^2}\; .
\end{equation}
These formulas are the archetype of the results obtained in this
paper.  We also derive similar formulas for bilocal operators of the
form $\big<\phi_a^*(0){\cal W}_{ab}(\gamma_{x0})\phi_b(x)\big>$, where
the point $x$ is separated from the origin by 1 or 2 lattice
spacings. In these operators, ${\cal W}_{ab}(\gamma_{x0})$ is a Wilson
line along a path $\gamma_{x0}$ connecting $x$ to 0, which is needed
in order to have a gauge invariant operator. We shall see that the
leading term of the expansion in powers of the lattice spacings does
not depend on the choice of this path, while the second term in
general depends on this choice.

In the rest of this paper, we use the lattice worldline formalism in
order to demonstrate these formulas for the coefficients of the
expansion. We first obtain intermediate representations of these
coefficients in terms of sums over all the closed random walks on the
lattice, which relate their values to some combinatorial properties of
random walks.  These formulas can then be transformed into the
integral representations listed above, by using the 2-dimensional
combinatorial formulas of the appendix \ref{app:area-aniso}.

\subsection{Outline of the paper}
In the section \ref{sec:phi2}, we derive in detail the worldline
formulation of the expectation value $\big<\phi_a^*(0)\phi_a(0)\big>$
on a lattice with isotropic spacings, and its short distance
expansion. A subsection is devoted to the discussion of infrared
divergences and their manifestation in the worldline formalism. We
also introduce in this section the Borel transformation that turns the
combinatorial sums into integrals.  In the section \ref{sec:bilocal},
we extend this study to bilocal operators, i.e. operators that contain
a $\phi^*$ and a $\phi$ evaluated at separate lattice spacings. This
extension is of great practical importance, because these operators
appear in the discretization of covariant derivatives.  The section
\ref{sec:aniso} generalizes all the previous results to a more general
lattice setup, where each direction of space-time has its own lattice
spacing. As an illustration, we study the limit where one of the
lattice spacings is much smaller than the others, and we apply this to
a discussion of the energy-momentum tensor. The section
\ref{sec:conclusions} is devoted to concluding remarks.

A number of more technical aspects are discussed in several
appendices. In the appendix \ref{app:Pn}, we show how this formalism
is modified on a finite lattice with periodic boundary conditions (in
the main part of the paper, we take the limit of zero lattice spacing
at fixed physical volume, so that the size of the lattice becomes
infinite and the boundary conditions are irrelevant). In the appendix
\ref{app:NL-D0}, we derive the leading coefficient of the expansion
for bilocal operators with an arbitrary separation between the two
fields. The appendix \ref{app:area-aniso} discusses all the
combinatorial formulas that are necessary in the case of anisotropic
lattices, and in \ref{app:mathieu} we recall the connection between
the statistics of the areas of closed loops on a two-dimensional
lattice and the spectral properties of the so-called almost-Mathieu
operator. In the appendix \ref{app:LPT}, we obtain the short distance
expansion of $\big<\phi_a^*(0)\phi_a(0)\big>$ from lattice
perturbation theory, mainly to illustrate the technical complexity of
this approach. In the appendix \ref{app:cont-time}, we study from the
outset a hybrid description in which one of the directions (e.g. time)
is treated as a continuous variable, while the others remain
discretized and we show that this is equivalent to starting from a
fully discrete description and taking one lattice spacing to zero.

\section{Local operator $\big<\phi_a^*(0)\phi_a(0)\big>$}
\label{sec:phi2}
\subsection{Discrete heat kernel}
In order to establish the formalism, consider first the expectation
value of the operator $\big<\phi_a^*(0)\phi_a(0)\big>$ (the color
indices are summed to ensure gauge invariance) at one loop. To lighten
the notations, we evaluate the expectation value at the point
$x^\mu=0$, but all our considerations are completely general. Since
this expectation value is given by a 1-loop graph, we can first write
\begin{equation}
\big<\phi_a^*(0)\phi_a(0)\big>
=\big<x^\mu=0\big|
\frac{1}{D^2}
\big|x^\mu=0\big>\; .
\label{eq:exp-phi2}
\end{equation}
The standard heat kernel approach would be to write
\begin{equation}
\frac{1}{D^2}
=
\int_0^\infty ds\;\exp(-s\,D^2)\; .
\end{equation}
However, in our case it is more convenient to use a discrete version
of this formula\footnote{Up to a rescaling, the integer $n$ is a
  discrete version of the integration variable $s$.},
\begin{equation}
\frac{2d}{a^2D^2}
=
\sum_{n=0}^\infty (1-a^2D^2/2d)^n\; ,
\end{equation}
which is also exact.  We have multiplied $D^2$ by the lattice spacing
squared in order to get a dimensionless combination. The purpose of
the factor $2d$ (where $d$ is the number of space-time dimensions)
will become clearer later on.

Consider now a sequence of functions $P_n(x)$ defined on the lattice,
and satisfying the following iteration rule:
\begin{equation}
P_{n+1}=(1-a^2D^2/2d)\,P_n\; .
\label{eq:Iter}
\end{equation}
If we define $P_0(x)=\delta_{x}$, then we have
\begin{equation}
\big<0\big|
(1-a^2D^2/2d)^n
\big|0\big>=a^{-d}\;P_n(x) \; .
\end{equation}
If we interpret $P_0(x)$ as a probability distribution localized at
the point $x^\mu=0$, then $P_n$ is the probability distribution after
$n$ iterations of the process described in eq.~(\ref{eq:Iter}). In
other words, it is the probability that this process starts and
returns at the point $x^\mu=0$ after exactly $n$ steps.

\subsection{Vacuum  case}
Eq.~(\ref{eq:Iter}) may be rewritten as
\begin{equation}
P_{n+1}-P_n=-\frac{a^2}{2d}D^2 P_n\; .
\label{eq:diff}
\end{equation}
If we view the index $n$ as a discrete fictitious time, and if the
metric is Euclidean, then this is a discrete diffusion equation and
the evolution of the probability distribution $P_n$ can be remapped in
terms of random walks.

For illustration purposes, consider first the free case. We have
\begin{equation}
-\frac{a^2}{2d}D^2 f(i,\cdots) =-f(i,\cdots)+ \frac{f(i+1,\cdots)+f(i-1,\cdots)}{2d}+\cdots\; .
\end{equation}
The eq.~(\ref{eq:diff}) can then be written more explicitly as
\begin{equation}
P_{n+1}(i,\cdots)=\frac{P_n(i+1,\cdots)+P_n(i-1,\cdots)}{2d}+\cdots\; ,
\label{eq:Iter-1}
\end{equation}
where the sum in the numerator extends to all the nearest neighbors.
This equation describes a random walk where at each step one moves to
one of the adjacent sites of the lattice with probability $1/2d$.  We
see now the reason for the peculiar
normalization\footnote{Alternatively, one could view this
  normalization as choosing a specific ratio between the size of the
  steps in the fictitious time and the lattice spacing.} in
eq.~(\ref{eq:exp-phi2})~: by doing this, we can eliminate the term
proportional to $P_n(i,\cdots)$ in the right hand side, i.e. the possibility
for the random walk process to stall during the step\footnote{By
  excluding the possibility that the random walk stalls, we ensure
  that the number of steps $n$ is also the length of the path.}.

\begin{figure}[htbp]
\begin{center}
\resizebox*{6cm}{!}{\includegraphics{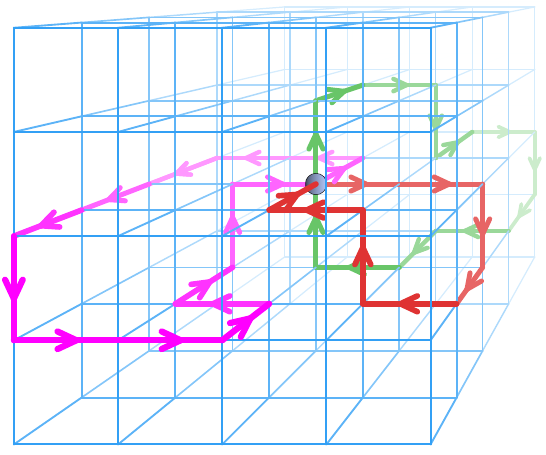}}
\end{center}
\caption{\label{fig:worldlines}Example of closed random paths on a
  3-dimensional cubic lattice. The blob indicates the location of the
  base point $0$.}
\end{figure}

$P_n(0)$ is the probability that such a random walk returns at the
point $0$ after $n$ steps. Geometrically, this means that the random
walk is a closed loop of length $n$. Since at each step, there are two
possibilities to move in each direction, the total number of random
walks of length $n$ is $(2d)^n$. $P_n(x)$ is thus the number of {\sl
  closed} random walks of length $n$, divided by the total number
$(2d)^n$.  Therefore, we can write
\begin{equation}
\big<\phi_a^*(0)\phi_a(0)\big>
=\frac{1}{2d a^{d-2}}\,\sum_{n=0}^\infty
\frac{1}{(2d)^{2n}}\;
\sum_{\gamma\in\Gamma_{2n}(0,0)} {\rm tr}_{\rm adj}\,\big(1\big)\; ,
\label{eq:exp-phi2-1}
\end{equation}
where $\Gamma_{2n}(0,0)$ is the set of all closed random walks of base
point $0$ (i.e. starting and ending at $0$) and length $2n$ on the
lattice (the length of such a closed path must be an even number). A
few of the closed paths involved in eq.~(\ref{eq:exp-phi2-1}) are
illustrated in the figure \ref{fig:worldlines}.

In the vacuum, the double sum is
independent of the lattice spacing. It is just a pure number that sets
the normalization of the result. The trace in the adjoint
representation comes from the summation over the color indices, and
brings a factor $N_c^2-1$.

\subsection{Non-zero background field}
On the lattice, the background field is represented in terms of
compact link variables $U_i(x)$ in order to preserve an exact gauge
invariance despite the discretization. In terms of these link
variables, the covariant derivative squared becomes,
\begin{eqnarray}
&&-\frac{a^2}{2d}D^2 f(i,\cdots) =-f(i,\cdots)
\nonumber\\
&&\qquad
+ \frac{U_1(i,\cdots)f(i+1,\cdots)+U_1^{-1}(i-1,\cdots)f(i-1,\cdots)}{2d}+\cdots\; .
\end{eqnarray}
Therefore, when the links are not unity, the random walk is biased by
the background field. The end result is that eq.~(\ref{eq:exp-phi2-1})
is modified into
\begin{equation}
\big<\phi_a^*(0)\phi_a(0)\big>
=\frac{1}{2d a^{d-2}}\,\sum_{n=0}^\infty
\frac{1}{(2d)^{2n}}\;
\sum_{\gamma\in\Gamma_{2n}(0,0)} {\rm tr}_{\rm adj}\,\big({\cal W}(\gamma)\big)\; .
\label{eq:exp-phi2-2}
\end{equation}
In words, the $SU(N_c)$ identity matrix in eq.~(\ref{eq:exp-phi2-1})
is replaced by a Wilson loop ${\cal W}(\gamma)$ obtained by
multiplying all the link variables along the closed contour
$\gamma$. This formula is manifestly gauge invariant, since
Wilson loops are gauge invariant.  Note also that this formula is exact at
1-loop on the lattice.

\subsection{Notations and basic facts about closed random walks}
In the previous subsections, we have introduced $\Gamma_{2n}(0,0)$,
the set of all the paths of length $2n$ drawn on the lattice, with
endpoints $0$ and $0$ (i.e. closed paths). More generally, we will
denote $\Gamma_{n}(0,x)$ the set of paths of length $n$ from $0$ to a
point $x$ (all these paths have the same parity, which is also the
parity of the sum of the coordinates of the point $x$). To avoid
encumbering the notation, we do not specify the dimension of the
lattice in which these paths should be considered, since the context
of the formula in which the notation appears is sufficient to make
this obvious.

When we need an explicit representation for a path, we denote it by
the sequence of the hops it contains, such as
\begin{equation}
\gamma= \hat{2}\;\hat{4}\;\hat{3}^{-1}\cdots\; ,
\end{equation}
(read from left to right.) The notation $\hat{3}^{-1}$ denotes a hop
in the $-x_3$ direction. The empty path will be denoted $\gamma={\bs
  1}$, and the concatenation of two paths $\gamma_1$ and $\gamma_2$ is
denoted by $\gamma=\gamma_1\otimes\gamma_2$ (read again from left to
right, so that $\gamma_1$ is the first part of the resulting
path). Obviously, ${\bs 1}\otimes\gamma=\gamma\otimes{\bs 1}=\gamma$.

In two dimensions, we will also introduce later in the paper the
subset ${\bs\Gamma}_{n_1,n_2}(0,x)$, made of all the paths
connecting $0$ to $x$ and making exactly $n_1$ hops in the $+x_1$
direction and $n_2$ hops in the $+x_2$ direction. The numbers
$n'_{1,2}$ of hops in the opposite directions, $-x_1$ and $-x_2$, do
not need to be specified explicitly since it can be inferred from
$n_{1,2}$ and the coordinates of the point $x$. Indeed, if $x=x_1
\hat{1}+x_2\hat{2}$, we have
\begin{equation}
n'_1=n_1-x_1\quad,\qquad n'_2=n_2-x_2\; .
\end{equation}
Obviously, $\Gamma_{n}(0,x)$ and ${\bs\Gamma}_{n_1,n_2}(0,x)$ are
related by
\begin{equation}
\Gamma_{n}(0,x)
=\bigcup_{n_1+n_2=\tfrac{n+x_1+x_2}{2}}{\bs\Gamma}_{n_1,n_2}(0,x)\; .
\end{equation}

\begin{figure}[htbp]
\begin{center}
\resizebox*{7cm}{!}{\includegraphics{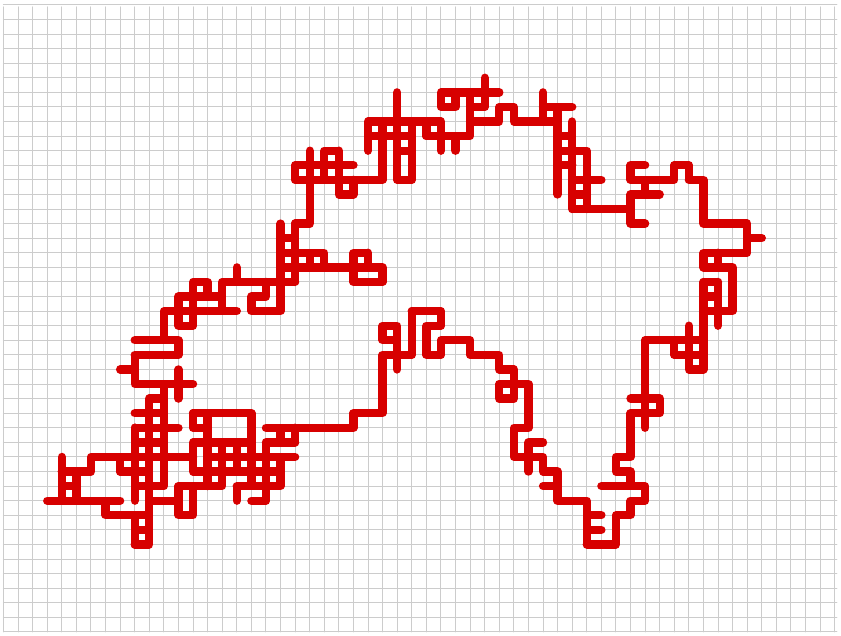}}
\end{center}
\caption{\label{fig:rw}Closed random walk on a square lattice in two
  dimensions. The length of the path is $2na$, while the diameter of
  the domain explored by the random walk is of order $\sqrt{n}a$ and
  its area is of order $na^2$.}
\end{figure}

In order to develop some intuition with formulas such as
eq.~(\ref{eq:exp-phi2-2}), let us recall here some elementary
properties of closed random walks. Let us consider a closed random
walk made of $2n$ hops, all illustrated in the figure \ref{fig:rw} in
two dimensions.
For such a random walk, one has the following properties:
\begin{itemize}
\item[{\bf i.}] the length of the path is obviously $2na$,
\item[{\bf ii.}] the typical size of the domain explored by the random walk grows only like $\sqrt{n}a$,
\item[{\bf iii.}] the area enclosed by the random walk (or the area of its projection on a plane in $d>2$ dimensions) grows as $na^2$.
\end{itemize}
The property {\bf ii} plays a role in the infrared behavior of the
quantity under consideration. Indeed, as we shall see later, infrared
singularities arise when the contribution of ``large'' random walks
does not decrease fast enough. Similarly, {\bf iii} plays a role in
the second term in the expansion in powers of the lattice spacing.

\subsection{Continuum limit $a\to 0$}
If we let the lattice spacing $a$ go to zero, while the background is
held fixed in physical units, the Wilson loop ${\cal W}(\gamma)$ goes
to the identity because the closed loop $\gamma$ shrinks to a tiny
loop of base point $0$. Therefore, it can be approximated by the
exponential of the magnetic flux across a surface $\Sigma$ whose
boundary is $\gamma$ (this surface is a tiling of elementary lattice
squares),
\begin{equation}
{\cal W}(\gamma)
\empile{\approx}\over{a\to 0}
\exp\Big\{ iga^2\sum_{\mu<\nu}A_{\mu\nu}(\gamma) F^{\mu\nu}_a(0)t^a\Big\}\; ,
\end{equation}
where $A_{\mu\nu}(\gamma)$ is the algebraic area, measured as a number
of plaquettes since we have already pulled out a factor $a^2$, of the
domain enclosed by the projection of the contour $\gamma$ on the
$(\mu,\nu)$ plane. The orientation of $\gamma$ dictates the
orientation of the projection, which in turn controls the sign of
$A_{\mu\nu}(\gamma)$.  Note that in this limit, the field strength can
be considered uniform across the entire lattice, and therefore
$F^{\mu\nu}$ is evaluated at the point $x^\mu=0$.

Since the loop size tends to zero when $a\to 0$, we can do a Taylor
expansion of the exponential. In order to get a non-trivial answer after
taking the trace, we must go to second order:
\begin{equation}
{\rm tr}_{\rm adj}\,\big({\cal W}(\gamma)\big)
\approx
{\rm tr}_{\rm adj}\,\big(1\big)
-\frac{g^2 a^4}{4}
\Big\{\sum_{\mu<\nu} A_{\mu\nu}(\gamma)F^{\mu\nu}_a(0)\Big\}
\Big\{\sum_{\rho<\sigma}A_{\rho\sigma}(\gamma) F^{\rho\sigma}_a(0)\Big\}
\end{equation}
(A factor $1/2$ comes from ${\rm tr}_{\rm adj}(t^a
t^b)=\tfrac{1}{2}\delta^{ab}$.)  By plugging this in the formula
(\ref{eq:exp-phi2-2}), we obtain the following expansion
\begin{equation}
\big<\phi_a^*(0)\phi_a(0)\big>
\approx\frac{1}{2d\, a^{d-2}}\left[{\bs C}_0\;{\rm tr}_{\rm adj}\big(1\big)
\!-\!
\frac{g^2a^4}{4}\,\sum_{\ontop{\mu<\nu}{\rho<\sigma}}F^{\mu\nu}_a(0)F^{\rho\sigma}_a(0)
\,{\bs C}_4^{\mu\nu;\rho\sigma}\right]
\label{eq:cont-exp}
\end{equation}
where the coefficients ${\bs C}_0$ and ${\bs C}_4^{\mu\nu;\rho\sigma}$ are purely
geometrical quantities defined by sums over all the closed loops on
the lattice
\begin{eqnarray}
{\bs C}_0&\equiv& \sum_{n=0}^\infty
\frac{1}{(2d)^{2n}}\;
\sum_{\gamma\in\Gamma_{2n}(0,0)} 1
\nonumber\\
{\bs C}_4^{\mu\nu;\rho\sigma}&\equiv&\sum_{n=0}^\infty
\frac{1}{(2d)^{2n}}\;
\sum_{\gamma\in\Gamma_{2n}(0,0)} A_{\mu\nu}(\gamma)A_{\rho\sigma}(\gamma)\; .
\label{eq:coeffs}
\end{eqnarray}
(Note that in these formulas, we have replaced $n\to 2n$ since only
random paths of even length can be closed.)  The equation
(\ref{eq:cont-exp}) provides an explicitly gauge invariant expansion
in powers of the lattice spacing. The coefficients (that remain to be
calculated) are geometrical quantities that depend on the dimension
and the lattice under consideration, but not on the background
field. 

In the second of eqs.~(\ref{eq:coeffs}), $A_{\mu\nu}(\gamma)$ is the area
of the surface enclosed by $\gamma$ projected on the $\mu\nu$
plane. Several remarks are in order about this quantity:
\begin{itemize}
\item These areas are ``algebraic'', in the sense that they may have a
  sign that takes into account the orientation of the boundary, and a
  multiplicity that depends on the winding number.
\item There are many surfaces with the same boundary $\gamma$. 
  $A_{\mu\nu}(\gamma)$ does not depend on this choice but only on the
  boundary.
\item They do not depend on the base point $x^\mu=0$. Specifying a
  base point is only necessary in order to avoid counting multiple
  times loops that have the same shape up to a translation.
\end{itemize}

\subsection{Zeroth order coefficient}
\label{sec:borel-C0}
It is possible to provide an integral expression for the coefficient
${\bs C}_0$, starting from the combinatorial formula that explicitly
counts the number of closed random walks in terms of the number of
hops in the $d$ directions (respectively $2n_1,2n_2,\cdots,2n_d$),
\begin{equation}
{\bs C}_0=
\sum_{n=0}^\infty\frac{(2n)!}{(2d)^{2n}}
\sum_{n_1+\cdots+n_d=n}
\frac{1}{n_1!^2\cdots n_d!^2}\; .
\end{equation}
The factor $(2n)!$ prevents the complete separation of the sums over
the $n_i$. However, it can be removed by a Borel trans\-for\-ma\-tion~:
\begin{equation}
{\bs C}_0=
\int_0^\infty dt\;e^{-t}\,A_d(\tfrac{t}{2d})\; ,
\end{equation}
where we denote
\begin{eqnarray}
A_d(x)&\equiv& \sum_{n=0}^\infty x^{2n}
\sum_{n_1+\cdots+n_d=n}
\frac{1}{n_1!^2\cdots n_d!^2}
\nonumber\\
&=&\left[\sum_{p=0}^\infty \frac{x^{2p}}{p!^2}\right]^d=I_0^d(2x)\; ,
\end{eqnarray}
where $I_0$ is a modified Bessel function of the first
kind. Therefore, we have
\begin{equation}
{\bs C}_0=
\int_0^\infty dt\;e^{-t}\,I_0^d(\tfrac{t}{d})\; .
\label{eq:C0-bessel}
\end{equation}
In 3 dimensions, this leads to an explicit formula (\cite{GradsR1}--\S6.612.6)~:
\begin{equation}
{\bs C}_0\empile{=}\over{d=3}
\frac{\sqrt{3}-1}{32\pi^3}\,\Gamma^2\big(\tfrac{1}{24}\big)\,
\Gamma^2\big(\tfrac{11}{24}\big)\approx 1.51638606\; ,
\end{equation}
while in 4 dimensions we have only been able to evaluate it
numerically,
\begin{equation}
{\bs C}_0\empile{\approx}\over{d=4}
1.23946712\; .
\end{equation}

\subsection{Variance of the areas of closed random walks}
\label{sec:A2-variance}
In the second term of the expansion in powers of the lattice spacing,
we need the quantity
\begin{eqnarray}
{\bs C}_4^{\mu\nu;\rho\sigma}&\equiv&\sum_{n=0}^\infty
\frac{1}{(2d)^{2n}}\;
\sum_{\gamma\in\Gamma_{2n}(0,0)} A_{\mu\nu}(\gamma)A_{\rho\sigma}(\gamma)\; .
\label{eq:coeff-A2}
\end{eqnarray}
A central result for the rest of our discussion in the isotropic case
is the value of this sum in two dimensions. In $d=2$, the variance of
the algebraic areas enclosed by closed random walks of length $2n$ is
given by the following formula~\cite{MingoN1} (eqs.~(1.4)--(1.5))~:
\begin{equation}
\left[\sum_{\gamma\in\Gamma_{2n}(0,0)} \left(A_{12}(\gamma)\right)^2\right]_{{\rm\ dim\ }2}
=
\left(\ontop{2n}{n}\right)^2\;\frac{n^2(n-1)}{6(2n-1)}\; .
\label{eq:A2-d2}
\end{equation}
This formula is all we need in order to evaluate the coefficient ${\bs
  C}_4$ for $d=2$. But note that this coefficient diverges in $d=2$:
using Stirling's asymptotic formula for the factorial, one can see
that the sum over the length $2n$ of the path is divergent (in this
formalism, this is the counterpart of an infrared divergent loop
integral in low dimension).

\begin{figure}[htbp]
\begin{center}
\resizebox*{5cm}{!}{\includegraphics{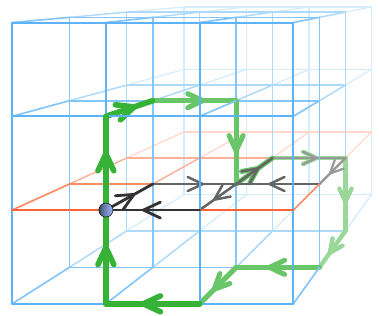}}
\end{center}
\caption{\label{fig:loop}Example of closed random walk on a
  3-dimensional cubic lattice, with $2n=8$ hops in the $12$ plane and
  $2m=4$ hops in the $3$-rd direction. The $12$ plane is highlighted in
  orange, and the projection of the closed loop on the plane is shown
  in gray (it has an area $A_{12}=-2$ in this example). The blob indicates
  the location of the base point $0$.}
\end{figure}

In higher dimensions, the first thing to notice is that it is
sufficient to consider $\mu=\rho,\nu=\sigma$ (if there is a mismatch
of the indices, the average over all closed loops gives zero since the
area is signed). For the sake of definiteness, let us choose
$\mu=1,\nu=2$. For a given closed loop $\gamma$, the area
$A_{12}(\gamma)$ is the area of its projection on the $12$
plane. Every closed random walk in $d$ dimensions can be decomposed
into hops that are in the $12$ plane, and hops orthogonal to this
plane (see the figure \ref{fig:loop}).
The latter disappear in the projection on the $12$ plane, and
therefore do not play any role in the calculation of the area
$A_{12}(\gamma)$. Moreover, the projection of $\gamma$ in the $12$
plane is itself a closed loop in $2$ dimensions, while the sequence of
the transverse hops is a closed loop in $d-2$ dimensions. Let us
denote $2n$ the number of hops in the $12$ plane and $2m$ the number
of hops in the transverse directions. One can rewrite the coefficient
${\bs C}_4$ as follows
\begin{eqnarray}
{\bs C}_4^{12;12}&\equiv&
\sum_{m,n=0}^\infty
\frac{1}{(2d)^{2(n+m)}}\;\left(\ontop{2(m+n)}{2m}\right)\nonumber\\
&&\quad\times
\left[\sum_{\sigma\in\Gamma_{2m}(0,0)} \!\!\!1\right]_{{\rm dim\ }d-2}\times
\left[\sum_{\gamma\in\Gamma_{2n}(0,0)} \!\!\!\left(A_{12}(\gamma)\right)^2\right]_{{\rm dim\ }2}\; .
\label{eq:coeff-C4-1}
\end{eqnarray}
The binomial factor in the first line counts the number of ways to
intertwine the $2m$ transverse hops and the $2n$ in-plane hops. In the
second line, the first factor is the number of length $2m$ closed
random walks in $d-2$ dimensions, and the second factor is the squared
area summed over all closed loops of length $2n$ in two dimensions.
This latter factor is given by eq.~(\ref{eq:A2-d2}).  For the first
factor, since we are interested primarily in $d=3$ ($d-2=1$) and $d=4$
($d-2=2$), we can use the following standard results
\begin{eqnarray}
&&
\left[\sum_{\sigma\in\Gamma_{2m}(0,0)} \!1\right]_{{\rm dim\ }1} = \left(\ontop{2m}{m}\right)
\nonumber\\
&&
\left[\sum_{\sigma\in\Gamma_{2m}(0,0)} \!1\right]_{{\rm dim\ }2} = \left(\ontop{2m}{m}\right)^2\; .
\end{eqnarray}
Therefore, for these dimensions, we find the following expressions for
the coefficient ${\bs C}_4^{12;12}$,
\begin{equation}
{\bs C}_4^{12;12}
\empile{=}\over{d=3}
\sum_{l=0}^\infty\frac{(2l)!}{6^{2l}}\sum_{n=0}^l\frac{(2n)!}{(l-n)!^2n!^4}\frac{n^2(n-1)}{6(2n-1)}\; ,
\label{eq:C4-3d}
\end{equation}
and
\begin{equation}
{\bs C}_4^{12;12}
\empile{=}\over{d=4}
\sum_{l=0}^\infty\frac{(2l)!}{8^{2l}}\sum_{n=0}^l
\frac{(2(l-n))!(2n)!}{(l-n)!^4n!^4}\frac{n^2(n-1)}{6(2n-1)}\; .
\label{eq:C4-4d}
\end{equation}
For a given total length $2l=2m+2n$ of the random walks, the variance
of the projected algebraic area is thus expressed as a sum of $l+1$
terms, whose evaluation is very easy (especially compared to a direct
evaluation by exhausting the list of all random walks of this length,
since there are $(2d)^{2l}$ such walks). In the table \ref{tab:3d}, we
list as an example the summands (at fixed $l$) for the 3 dimensional
case (eq.~(\ref{eq:C4-3d})). We have performed an exhaustive search of
all the closed random walks up to $2l=14$, and we have checked the
agreement between this direct computation and the formula
(\ref{eq:C4-3d}). The values listed for $2l>14$ were solely obtained
from eq.~(\ref{eq:C4-3d}).
\begin{table}[htbp]
\begin{center}
\begin{tabular}{c|c|c|c}
$2l$ & \#(paths) &\#(closed paths) & $\tfrac{\sum_{\gamma\in\Gamma_{2l}(0,0)}(A_{12}(\gamma))^2}{6^{2l}}$\\{\ }&&&\\
\hline
2& 36  & 6 & 0.0000000000\\
4& 1296&90 & 0.0061728395\\
6& 46656&1860&0.0102880658\\
8&1679616&44730&0.0133363816\\
10&60466176&1172556&0.0158369532\\
12&2176782336&32496156&0.0180064306\\
14&78364164096&936369720&       0.0199490385\\
\hline
20&$36^{10}$&& 0.0249038527\\
100&$36^{50}$&& 0.0600254031\\
200&$36^{100}$&& 0.0856471034\\
1000&$36^{500}$&& 0.1928668060\\
2000&$36^{1000}$&& 0.2729940025\\
\end{tabular}
\end{center}
\caption{\label{tab:3d}Exhaustive enumeration of random walks and
  closed random walks in 3 dimensions, up to the length $2l=14$. The
  last column gives the corresponding contribution to ${\bs
    C}_4^{12;12}$. The values for $2l>14$ are obtained from
  eq.~(\ref{eq:C4-3d}).}
\end{table}

It is also possible to establish an integral representation of ${\bs
  C}_4^{12;12}$, valid in any dimension $d$, similar to
eq.~(\ref{eq:C0-bessel}) for ${\bs C}_0$. The first step is to
introduce the combinatorial representation for the factor that counts
the closed random walks of length $2m$ in $d-2$ dimensions. This leads
to
\begin{eqnarray}
{\bs C}_4^{12;12}
&=&
\sum_{m,n=0}^\infty
\frac{1}{(2d)^{2(m+n)}}
\frac{(2(m+n))!}{(2m)!(2n)!}
\sum_{m_1+\cdots+m_{d-2}=m}\frac{(2m)!}{m_1!^2\cdots m_{d-2}!^2}
\nonumber\\
&&\qquad\qquad\qquad\times\;
\frac{(2n)!^2}{n!^4}\frac{n^2(n-1)}{6(2n-1)}\; .
\end{eqnarray}
We can separate the sums over $m$ and $n$ by a Borel transformation,
\begin{equation}
{\bs C}_4^{12;12}
=
\int_0^\infty
dt\;e^{-t}\;C_d(\tfrac{t}{2d})\; ,
\end{equation}
with
\begin{eqnarray}
C_d(x)&\equiv& \sum_{m,n=0}^\infty
\frac{x^{2(m+n)}}{(2m)!(2n)!}
\sum_{m_1+\cdots+m_{d-2}=m}\frac{(2m)!}{m_1!^2\cdots m_{d-2}!^2}
\frac{(2n)!^2}{n!^4}\frac{n^2(n-1)}{6(2n-1)}
\nonumber\\
&=&\frac{x^2}{3}\;I_0^{d-2}(2x)\;I_1^2(2x)\; .
\end{eqnarray}
Therefore, we have
\begin{equation}
{\bs C}_4^{12;12}=\frac{1}{12d^2}
\int_0^\infty dt\; e^{-t}\;t^2\;I_0^{d-2}(\tfrac{t}{d})\;I_1^2(\tfrac{t}{d})\; .
\label{eq:C4-bessel}
\end{equation}

\subsection{Infrared divergences}
\label{sec:mass}
As one can see in the table \ref{tab:3d}, the summands do not decrease
at large path lengths $l$, and the sum over $l$ is divergent. In
$d=3$, the summand grows as $l^{1/2}$ for large path lengths, and it
goes to a constant in $d=4$. If we cut off the sum over $l$ at some
$l_{\rm max}$, this implies that
\begin{equation}
{\bs C}_4^{12;12}
\empile{\sim}\over{d=3}
l_{\rm max}^{3/2}\qquad,\qquad
{\bs C}_4^{12;12}
\empile{\sim}\over{d=4}
l_{\rm max}\; .
\label{eq:C4-IR-div}
\end{equation}
This divergence for random walks that explore large regions in
spacetime is the manifestation in the worldline formalism of an
infrared singularity. This is corroborated by the fact the divergence
is milder in $d=4$ compared to $d=3$.  In the integral representation
(\ref{eq:C4-bessel}), this singularity appears as a divergence of the
integral at large $t$~: using the fact that $I_n(t)\sim t^{-1/2}e^t$
at large $t$, we see that the exponential factors cancel and that the
remaining algebraic factors decrease fast enough for convergence only
if $d>6$.

In order to further investigate this, let us add a mass
term\footnote{One may view this mass as a temporary regulator for the
  infrared sector. Note that if we do not expand the observable in
  powers of the background field, then the background field itself
  would provide a natural infrared cutoff. The sum over $l$ would
  naturally be cutoff when random walks reach a size comparable to the
  coherence length of the background field. For instance, if the
  background field is incoherent beyond the length scale $Q^{-1}$,
  then values $l\gtrsim (Qa)^{-2}$ are suppressed.} to the Lagrangian
of the scalar field,
\begin{equation}
{\cal L}\equiv \sum_{\mu=1}^d(D_\mu \phi)^*(D_\mu\phi)-m^2\phi^*\phi\; .
\label{eq:L-mass}
\end{equation}
In order to arrive again at a sum of random walks without stalls, one
should start from the following formula
\begin{equation}
\frac{2\tilde{d}}{a^2(D^2+m^2)}
=
\sum_{n=0}^\infty (1-a^2\tfrac{D^2+m^2}{2\tilde{d}})^n\; ,
\end{equation}
where we have defined $\tilde{d}\equiv d+\frac{1}{2}m^2a^2$. Most of
the discussion is unchanged and eq.~(\ref{eq:exp-phi2-2}) becomes
\begin{equation}
\big<\phi_a^*(0)\phi_a(0)\big>
=\frac{1}{2\tilde{d} a^{d-2}}\,\sum_{n=0}^\infty
\frac{1}{(2\tilde{d})^n}\;
\sum_{\gamma\in\Gamma_n(0,0)} {\rm tr}_{\rm adj}\,\big({\cal W}(\gamma)\big)\; .
\label{eq:exp-phi2-2-mass}
\end{equation}
A crucial difference is that each hop in the random walk is now
weighted by a factor $1/2\tilde{d}$ instead of $1/2d$. Since
$\tilde{d}>d$, this leads to an exponential reduction of the
contribution of long random walks. The eqs.~(\ref{eq:C4-3d}) and
(\ref{eq:C4-4d}) are modified into
\begin{equation}
{\bs C}_4^{12;12}
\empile{=}\over{d=3}
\sum_{l=0}^\infty\frac{(2l)!}{(6+m^2a^2)^{2l}}\sum_{n=0}^l\frac{(2n)!}{(l-n)!^2n!^4}\frac{n^2(n-1)}{6(2n-1)}\; ,
\label{eq:C4-3d-mass}
\end{equation}
and
\begin{equation}
{\bs C}_4^{12;12}
\empile{=}\over{d=4}
\sum_{l=0}^\infty\frac{(2l)!}{(8+m^2a^2)^{2l}}\sum_{n=0}^l
\frac{(2(l-n))!(2n)!}{(l-n)!^4n!^4}\frac{n^2(n-1)}{6(2n-1)}\; .
\label{eq:C4-4d-mass}
\end{equation}
These discrete sums also have an integral representation in terms of
modified Bessel functions,
\begin{equation}
{\bs C}_4^{12;12}=\frac{1}{12\tilde{d}^2}
\int_0^\infty dt\; e^{-t}\;t^2\;I_0^{d-2}(\tfrac{t}{\tilde{d}})\;I_1^2(\tfrac{t}{\tilde{d}})\; .
\label{eq:C4-bessel-mass}
\end{equation}
The derivation of this formula follows the same line as that of eq.~(\ref{eq:C4-bessel}).

\begin{figure}[htbp]
\begin{center}
\resizebox*{9cm}{!}{\includegraphics{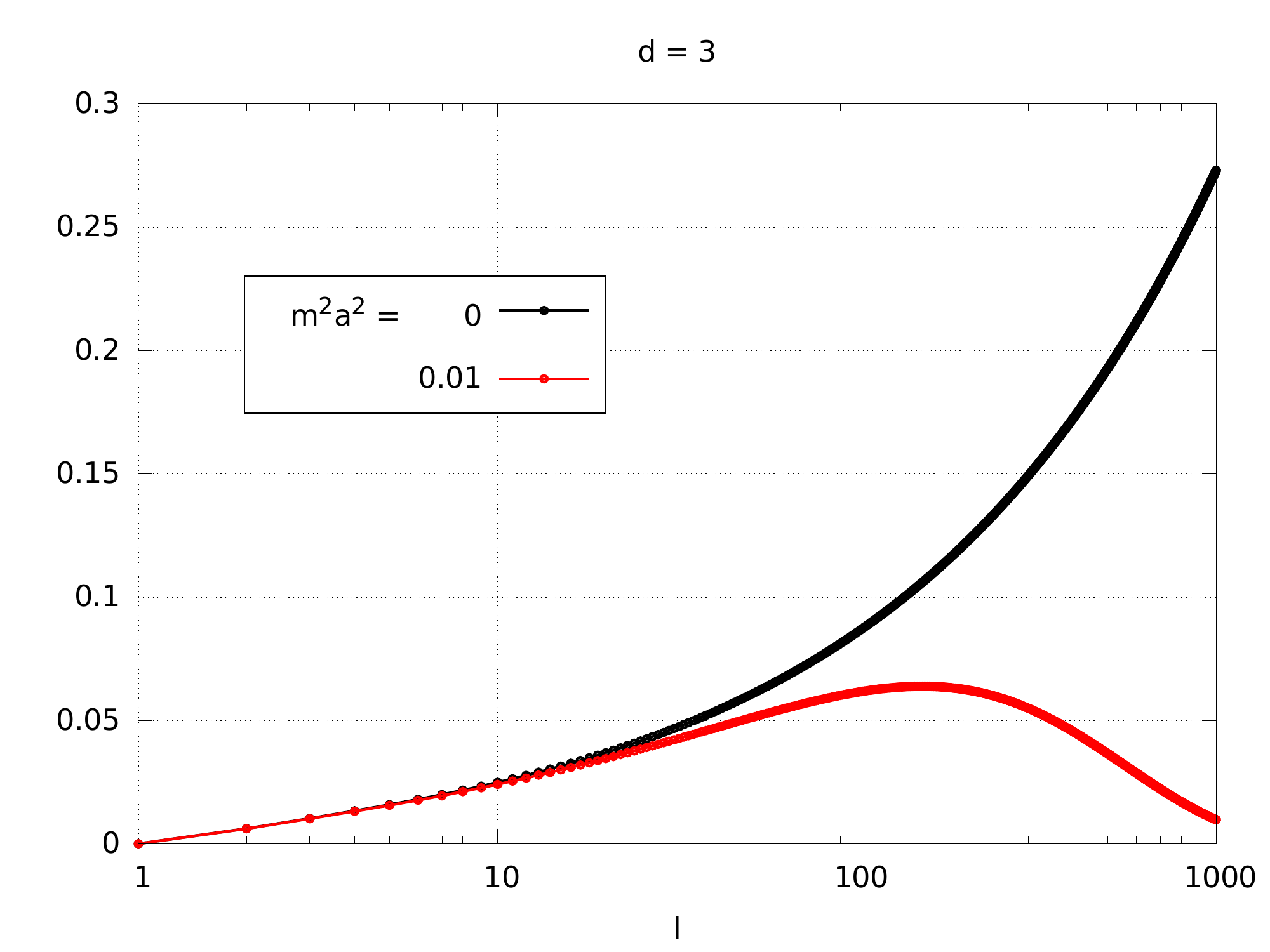}}
\end{center}
\caption{\label{fig:mass3d}Summand in $3$ dimensions, in the massless
  case (black curve) and for $m^2a^2=0.01$ (red curve).}
\end{figure}

The modification of the behavior for large random walks is readily
seen,
\begin{equation}
\frac{1}{(2\tilde{d})^{2l}}=\frac{1}{(2d)^{2l}}\frac{1}{(1+\tfrac{m^2a^2}{2d})^{2l}}
\approx
\frac{1}{(2d)^{2l}}\,e^{-lm^2a^2/d}\; .
\label{eq:IR-mass-impro}
\end{equation}
The regularizing effect of the mass becomes effective for lengths
$l\gtrsim d/m^2a^2$. This corresponds to random walks that explore a
domain of size $r\gtrsim \sqrt{d}/m$, i.e. of the order of the Compton
wavelength.  The values of the summand in $d=3$ are shown for
$m^2a^2=0.01$ in the figure \ref{fig:mass3d}, and compared to the
massless case.
With a non-zero mass, the summand increases until it reaches a maximum
and then decreases exponentially, which ensures the convergence of
the sum over $l$. Note that in doing this, we do not expand the $a$
dependence that comes from $2d+m^2a^2$. 

After this infrared regularization, the singular behaviors of
eq.~(\ref{eq:C4-IR-div}) would be replaced by
\begin{equation}
{\bs C}_4^{12;12}
\empile{\sim}\over{d=3}
(ma)^{-3}\qquad,\qquad
{\bs C}_4^{12;12}
\empile{\sim}\over{d=4}
(ma)^{-2}\; .
\label{eq:C4-IR-div-1}
\end{equation}
These formulas contain inverse powers of the lattice spacing. This
tells us that the expansion of eq.~(\ref{eq:cont-exp}), in which the
second term is suppressed by $a^4$, is upset by infrared
singularities. In fact, the second term does not vanish when $a\to 0$,
but instead is a term of order $a^0$, both in $d=3$ and $d=4$,
\begin{equation}
\big<\phi^*_a(0)\phi_a(0)\big>
\empile{=}\over{d=3,4} a^{2-d}\oplus m^{2-d}a^0\; .
\end{equation}
The same is true when the infrared regularization is provided by the
background field rather than by a mass. In this case, the mass is
replaced by the coherence scale $Q$ of the background field in the
above counting. Note that a proper calculation of the terms in $a^0$
requires a non-perturbative treatment of the background field (as we
have seen before, it was the expansion in powers of the background
field that caused the infrared divergence in the first place).

\section{Bilocal operators $\big<\phi_a^*(0){\cal W}_{ab}(\gamma_{x0})\phi_b(x)\big>$}
\label{sec:bilocal}
\subsection{Worldline representation}
Up to now, we have considered only the local operator
$\phi^*(0)\phi(0)$. However, since derivatives are represented on the
lattice as finite differences, we need also to consider composite
operators made of two elementary fields evaluated at different lattice
sites. Let us illustrate this by the discretization of the operator
$\phi^*D_\mu D_\mu\phi$. On the lattice, this can be represented as
follows,
\begin{eqnarray}
\phi^*_a(0)(D_\mu D_\mu\phi)_a(0)
&=&
\frac{1}{a^2}\Big[
\phi_a^*(0)U_\mu^{ab}(0)\phi_b(\widehat{\mu})
\nonumber\\
&&\quad+
\phi_a^*(0)U_\mu^{\dagger ab}(-\widehat{\mu})\phi_b(-\widehat{\mu})
-2\phi_a^*(0)\phi_a(0)\Big]\; .
\nonumber\\
&&
\label{eq:phi2-NL}
\end{eqnarray}
Here, one of the derivatives has been discretized as a forward
derivative and the other as a backward one. We have used covariant
derivatives for the operator to be gauge invariant. After
discretization, this leads to link variables connecting the two
lattice sites where the scalar field and its complex conjugate are
evaluated.

All the terms in the right hand side of eq.~(\ref{eq:phi2-NL}) belong
to a class of operators that contain two fields $\phi^*\cdots\phi$
linked by a Wilson line,
\begin{equation}
\big<\phi_a^*(0){\cal W}_{ab}(\gamma_{x0})\phi_b(x)\big>\; .
\end{equation}
In this equation, $\gamma_{x0}$ is a path (drawn on the edges of the
lattice) connecting the points $x$ and $0$ where the two fields are
evaluated. The Wilson line and the summation over the color indices
ensure that this operator is gauge invariant. However, the choice of
this path is arbitrary and is therefore a part of the definition of
the operator under consideration. Except when the background field is
a pure gauge, different paths correspond to operators that have
distinct expectation values.

\begin{figure}[htbp]
\begin{center}
\resizebox*{6cm}{!}{\includegraphics{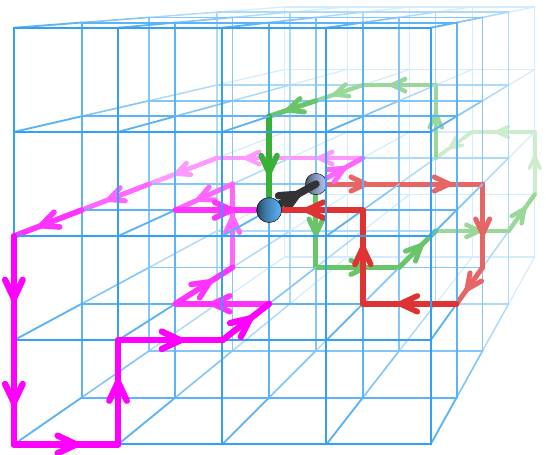}}
\end{center}
\caption{\label{fig:worldlines-offset}Example of random paths that
  appear in the worldline representation of bilocal operators, on a
  3-dimensional cubic lattice. The blobs indicate the locations of the
  base points $0$ and $x$ (here separated by 1 lattice spacing). The
  black link is the Wilson line ${\cal W}(\gamma_{x0})$ inserted
  between the two fields for gauge invariance.}
\end{figure}

It is quite straightforward to generalize the derivation done in the
section \ref{sec:phi2} in order to obtain the lattice worldline
representation for this type of operator,
\begin{equation}
\big<\phi_a^*(0){\cal W}_{ab}(\gamma_{x0})\phi_b(x)\big>
=\frac{1}{2d a^{d-2}}\,\sum_{n=0}^\infty
\frac{1}{(2d)^n}\;
\sum_{\gamma\in\Gamma_n(0,x)} {\rm tr}_{\rm adj}\,\big({\cal W}(\gamma_{x0}\otimes \gamma)\big)\; .
\label{eq:exp-phi2-NL}
\end{equation}
In this formula, $\Gamma_n(0,x)$ denotes the set of all the random
walks of length $n$ that start at the point $0$ and end at the point
$x$, and $\gamma_{x0}\otimes \gamma$ denotes the closed path obtained by
concatenating one of these paths and the contour $\gamma_{x0}$ used in
the definition of the operator, as illustrated in the figure
\ref{fig:worldlines-offset}.

\subsection{Continuum limit}
Let us now consider the continuum limit of these expectation
values. There are two non-equivalent ways to view this limit:
\begin{itemize}
\item[{\bf i.}] Keep the separation $x$--$0$ between the two points fixed
  in absolute units. When $a\to 0$, this interval becomes infinite in
  lattice units. This leads to milder ultraviolet divergences, since
  when the $a\to 0$ limit is performed in this way, one is in fact
  considering a non-local operator.
\item[{\bf ii.}] Keep $x$ and $0$ at fixed locations on the lattice
  while $a\to0$. Therefore, the spacing $x$--$0$ goes to zero in absolute
  units, and the limit $a\to 0$ corresponds to a local composite
  operator.
\end{itemize}
The limit {\bf ii} is the one we consider here, since we are
interested in operators such as the one in eq.~(\ref{eq:phi2-NL}). The
eq.~(\ref{eq:cont-exp}) becomes
\begin{eqnarray}
\big<\phi_a^*(0){\cal W}_{ab}(\gamma_{x0})\phi_b(x)\big>
&=&\frac{1}{2d a^{d-2}}\;\Big[{\bs D}_{0,\gamma_{x0}}\;{\rm tr}_{\rm adj}\,\big(1\big)
\nonumber\\
&&\,
-
\frac{g^2a^4}{4}\;\sum_{\ontop{\mu<\nu}{\rho<\sigma}}F^{\mu\nu}_a(x)F^{\rho\sigma}_a(x)
\;{\bs D}_{4,\gamma_{x0}}^{\mu\nu;\rho\sigma}+\cdots\Big]\; ,
\label{eq:cont-exp-NL}
\end{eqnarray}
where the coefficients ${\bs D}_{0,\gamma_{x0}}$ and ${\bs
  D}_{4,\gamma_{x0}}^{\mu\nu;\rho\sigma}$ are generalizations of the
coefficients ${\bs C}_{0}$ and ${\bs C}_{4}^{\mu\nu;\rho\sigma}$
introduced earlier
\begin{eqnarray}
{\bs D}_{0,\gamma_{x0}}&\equiv& \sum_{n=0}^\infty
\frac{1}{(2d)^{n}}\;
\sum_{\gamma\in\Gamma_{n}(0,x)} 1
\nonumber\\
{\bs D}_{4,\gamma_{x0}}^{\mu\nu;\rho\sigma}&\equiv&\sum_{n=0}^\infty
\frac{1}{(2d)^{n}}\;
\sum_{\gamma\in\Gamma_{n}(0,x)} A_{\mu\nu}(\gamma_{x0}\otimes \gamma)A_{\rho\sigma}(\gamma_{x0}\otimes \gamma)\; .
\label{eq:coeffs-NL}
\end{eqnarray}
The coefficient ${\bs D}_{0,\gamma_{x0}}$ depends only on the points
$0,x$ but not on the path $\gamma_{x0}$ chosen to connect
them. However, the coefficient ${\bs D}_{4,\gamma_{x0}}$ a priori
depends on the path $\gamma_{x0}$ as well, since the projected areas
$A_{\mu\nu}(\gamma_{x0}\otimes \gamma)$ depend on the shape of the
closed contour $\gamma_{x0}\otimes \gamma$.

\subsection{Bilocal operators with a 1-hop separation}
Let us first consider the specific case where the points $0$ and $x$
are nearest neighbors on the lattice and the path $\gamma_{x0}$ that
connects them is the shortest possible, i.e. an elementary link on the
cubic lattice. 
\subsubsection{Coefficient ${\bs D}_{0,\hat{x}}$}
For the coefficient ${\bs D}_{0,\gamma_{x0}}$, we can
without loss of generality assume that the link $\gamma_{x0}$ is a
link $\gamma_{x0}=\hat{x}$.  The random walks that connects $0$ to
$x=0+\hat{x}$ must have an odd length, $n=2m+1$. Simple combinatorics
leads to the following expression,
\begin{equation}
{\bs D}_{0,\hat{x}}=
\sum_{m=0}^\infty\frac{1}{(2d)^{2m+1}}\sum_{n_1+\cdots+n_d=m}\frac{(2m+1)!}{n_1!(1+n_1)!n_2!^2\cdots n_d!^2}\; .
\label{eq:D0-1}
\end{equation}
By a Borel transformation similar to the one used in the section
\ref{sec:borel-C0}, we can derive the following integral
representation for this coefficient\footnote{In the appendix
  \ref{app:NL-D0}, we derive an integral expression for this leading
  order coefficient ${\bs D}_{0,\gamma_{x0}}$ when $0$ and $x$ are not
  nearest neighbors.}
\begin{equation}
{\bs D}_{0,\hat{x}}=
\int_0^\infty dt\;e^{-t}\; I_1(\tfrac{t}{d})I_0^{d-1}(\tfrac{t}{d})\; .
\end{equation}
In 3 dimensions, this leads to a closed expression (\cite{GradsR1}--\S6.612.6),
\begin{equation}
{\bs D}_{0,\hat{x}}\empile{=}\over{d=3} \frac{\sqrt{3}-1}{32\pi^3}\,\Gamma^2\big(\tfrac{1}{24}\big)\,
\Gamma^2\big(\tfrac{11}{24}\big)-1\approx 0.51638606\; ,
\end{equation}
while in 4 dimensions we get
\begin{equation}
{\bs D}_{0,\hat{x}}\empile{\approx}\over{d=4}0.2394671218\; .
\label{eq:D0-4}
\end{equation}

\subsubsection{Coefficient ${\bs D}_{4,\hat{\sigma}}^{12;12}$}
We must extend the discussion of the subsection \ref{sec:A2-variance}
to a summation over all closed loops with a fixed link, of the type
shown in the figure \ref{fig:worldlines-offset}. We need to consider
two cases:
\begin{itemize}
\item[{\bf a.}] The link $\gamma_{x0}$ connecting the two points is not in
  the plane $\mu\nu$ on which we project the areas. The archetype of
  this case is $\mu\nu=12$ and $\gamma_{x0}=\hat{3}$ (one elementary link
  in the $+x_3$ direction).
\item[{\bf b.}] The link $\gamma_{x0}$ lies in the $\mu\nu$ plane. For
  instance, $\mu\nu=12$ and $\gamma_{x0}=\hat{1}$.
\end{itemize}

The first case {\bf a} is the simplest, because the fixed link does not
affect in any way the projected area in the $12$ plane.  The fixed
link $\gamma_{x0}=\hat{3}$ only alters the counting of the hops that are
orthogonal to the $12$ plane. We now have
\begin{eqnarray}
{\bs D}_{4,\hat{3}}^{12;12}&\equiv&
\sum_{m,n=0}^\infty
\frac{1}{(2d)^{2(n+m)+1}}\;\left(\ontop{2(m+n)+1}{2m+1}\right)\nonumber\\
&&\quad\times
\left[\sum_{\sigma\in\Gamma_{2m+1}(0,\hat{3})} \!\!\!1\right]_{{\rm dim\ }d-2}
\!\!\!\times
\left[\sum_{\gamma\in\Gamma_{2n}(0,0)} \!\!\!\left(A_{12}(\gamma)\right)^2\right]_{{\rm dim\ }2}\!\!.
\label{eq:coeff-D4-1}
\end{eqnarray}
Note that the number of hops in the transverse directions must be odd
(we denote it $2m+1$). We can still use the formula (\ref{eq:A2-d2})
for the variances of the areas in dimension 2 (the second factor on
the second line). The first factor on the second line counts the
number of random walks of length $2m+1$ in the transverse directions
that connect the points $0$ and $\hat{3}$. In $d=3$ and $d=4$, this factor
is given by
\begin{eqnarray}
&&
\left[\sum_{\sigma\in\Gamma_{2m+1}(0,\hat{3})} \!1\right]_{{\rm dim\ }1} = \left(\ontop{2m+1}{m}\right)
\nonumber\\
&&
\left[\sum_{\sigma\in\Gamma_{2m+1}(0,\hat{3})} \!1\right]_{{\rm dim\ }2} = 
\sum_{p+q=m}\frac{(2m+1)!}{p!(p+1)!q!^2}
=\left(\ontop{2m+1}{m}\right)^2\; .
\label{eq:RD-fixed-edge}
\end{eqnarray}
Therefore, we find the following expressions for the coefficient ${\bs
  D}_{4,\hat{3}}^{12;12}$,
\begin{equation}
{\bs D}_{4,\hat{3}}^{12;12}
\empile{=}\over{d=3}
\sum_{l=0}^\infty\frac{(2l+1)!}{6^{2l+1}}\sum_{n=0}^l\frac{(2n)!}{(l-n)!(l-n+1)!n!^4}\frac{n^2(n-1)}{6(2n-1)}\; ,
\label{eq:D4-z-3d}
\end{equation}
and
\begin{equation}
{\bs D}_{4,\hat{3}}^{12;12}
\empile{=}\over{d=4}
\sum_{l=0}^\infty\frac{(2l+1)!}{8^{2l+1}}\sum_{n=0}^l
\frac{(2(l-n)+1)!(2n)!}{(l-n)!^2(l-n+1)!^2n!^4}\frac{n^2(n-1)}{6(2n-1)}\; .
\label{eq:D4-z-4d}
\end{equation}

In the case {\bf b}, the formula (\ref{eq:A2-d2}) must be modified in
order to sum only over loops that start at the point $0$ and end at the
point $\hat{1}$. Let us consider random walks of length $2n-1$, so
that the length is $2n$ after adding a hop $\hat{1}^{-1}$ to return to the
starting point and close the loop. Consider the set of all the closed
random walks of length $2n$, over which the sum in
eq.~(\ref{eq:A2-d2}) is performed. After $2n-1$ steps, these walks
must be at one of the 4 four points $\hat{1}$, $-\hat{1}$, $\hat{2}$
or $-\hat{2}$ (since at the next hop they return at $0$).  Thus the
set $\Gamma_{2n}(0,0)$ of these closed random walks can be partitioned
into four subsets according to the point reached at the step
$2n-1$. These subsets are identical up to rotations of angles multiple
of $\pi/2$, and therefore each of these subsets has the same
contribution to the variance of the area (since the area is invariant
under these rotations). Since we are now interested only in the random
walks that reach the point $\hat{1}$ after $2n-1$ steps, we see that
eq.~(\ref{eq:A2-d2}) must be replaced by\footnote{In order to obtain
  this formula, one may also use eq.~(\ref{eq:area-nxy-0x}) from the
  appendix \ref{app:area-aniso} and sum over $1\le n_1\le
  n$,\begin{equation*}
    \sum_{n_1+n_2=n}\frac{(2(n_1+n_2)-1)!}{n_1!(n_1-1)!
      n_2!^2}\;\frac{n_1 n_2}{3}
    =\frac{1}{4}\left(\ontop{2n}{n}\right)^2\;\frac{n^2(n-1)}{6(2n-1)}\;
    .
  \end{equation*} }
\begin{equation}
\left[\sum_{\gamma\in\Gamma_{2n-1}(0,\hat{1})} \left(A_{12}(\gamma)\right)^2\right]_{{\rm\ dim\ }2}
=
\frac{1}{4}
\left(\ontop{2n}{n}\right)^2\;\frac{n^2(n-1)}{6(2n-1)}\; .
\label{eq:A2-d2-x}
\end{equation}
Note that the combinatorial prefactor can also be written as
\begin{equation}
\frac{1}{4}
\left(\ontop{2n}{n}\right)^2 = \left(\ontop{2n-1}{n}\right)^2\; ,
\end{equation}
which is the number of 2-dimensional random walks of length $2n-1$
from $0$ to $\hat{1}$ (see the second of
eqs.~(\ref{eq:RD-fixed-edge})).  From this, it is straightforward to
find the analogue of eqs.~(\ref{eq:D4-z-3d}) and (\ref{eq:D4-z-4d}) in
the case where $\gamma_{x0}=\hat{1}^{-1}$~:
\begin{equation}
{\bs D}_{4,\hat{1}}^{12;12}
\empile{=}\over{d=3}
\sum_{l=0}^\infty\frac{(2l+1)!}{6^{2l+1}}\sum_{n=0}^l\frac{(2n+1)!}{(l-n)!^2n!^2(n+1)!^2}\frac{(n+1)^2n}{6(2n+1)}\; ,
\label{eq:D4-x-3d}
\end{equation}
and
\begin{equation}
{\bs D}_{4,\hat{1}}^{12;12}
\empile{=}\over{d=4}
\sum_{l=0}^\infty\frac{(2l+1)!}{8^{2l+1}}\sum_{n=0}^l
\frac{(2(l-n))!(2n+1)!}{(l-n)!^4n!^2(n+1)!^2}\frac{(n+1)^2n}{6(2n+1)}\; .
\label{eq:D4-x-4d}
\end{equation}
(Note that, compared to eq.~(\ref{eq:A2-d2-x}), we have shifted $n\to
n+1$, so that the number of hops in the $12$ plane is $2n+1$ instead
of $2n-1$.)

The combinatorial formulas (\ref{eq:D4-z-3d}), (\ref{eq:D4-z-4d}),
(\ref{eq:D4-x-3d}) and (\ref{eq:D4-x-4d}) can be transformed into an
integral by means of a Borel transformation. We find the following
results, valid in $d$ dimensions~:
\begin{equation}
{\bs D}_{4,\hat{3}}^{12;12}=\frac{1}{12d^2}\int_0^\infty dt\; e^{-t}\; t^2\; I_0^{d-3}(\tfrac{t}{d})I_1^3(\tfrac{t}{d})\; ,
\label{eq:D4z-bessel}
\end{equation}
and
\begin{equation}
{\bs D}_{4,\hat{1}}^{12;12}=\frac{1}{12d^2}\int_0^\infty dt\; e^{-t}\; t^2\; I_0^{d-1}(\tfrac{t}{d})I_1(\tfrac{t}{d})\; .
\label{eq:D4x-bessel}
\end{equation}

\subsubsection{Behavior for long random walks}
\begin{figure}[htbp]
\begin{center}
\resizebox*{9cm}{!}{\includegraphics{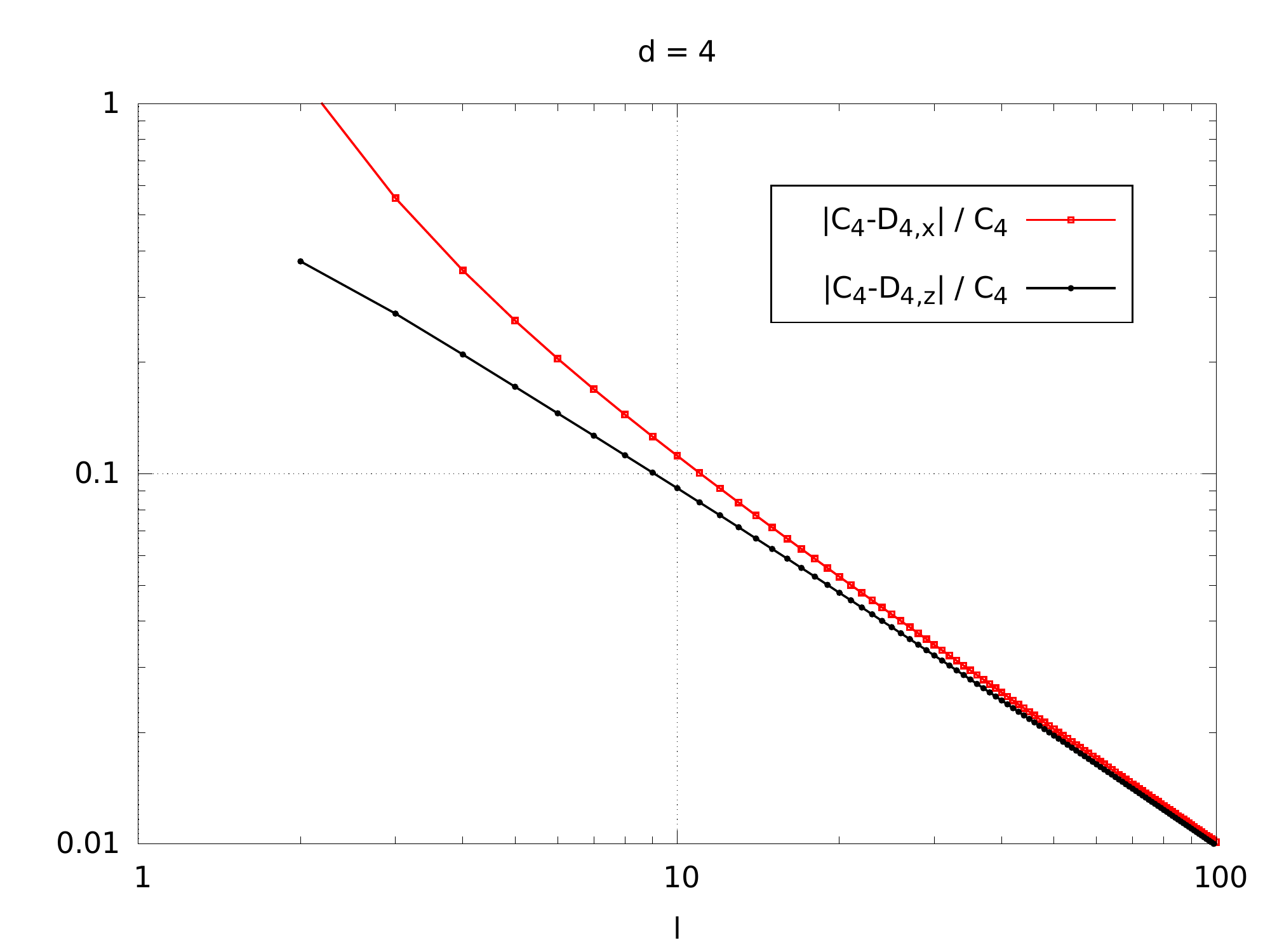}}
\end{center}
\caption{\label{fig:IR-diff}Relative difference between the summands
  in the coefficients ${\bs C}_4^{12,12}$, ${\bs
    D}_{4,\hat{1}}^{12,12}$ and ${\bs D}_{4,\hat{3}}^{12,12}$, as a
  function of the length of the random walk, in $d=4$ dimensions.}
\end{figure}

Having in mind the infrared divergences that manifest themselves in
the behavior of the summand in ${\bs C}_4^{12,12}$ for large path
lengths $2l$, it is interesting to study also the summand in the newly
introduced coefficients ${\bs D}_{4,\hat{1}}^{12,12}$ and ${\bs
  D}_{4,\hat{3}}^{12,12}$. In the figure \ref{fig:IR-diff}, we plot as
a function of the index $l$ the relative differences between these
quantities.
It appears that these differences decrease as $l^{-1}$ when $l\to
\infty$. In other words, the coefficients ${\bs C}_4^{12,12}$, ${\bs
  D}_{4,\hat{1}}^{12,12}$ and ${\bs D}_{4,\hat{3}}^{12,12}$ all have
the same behavior for large random walks. This is in fact quite
intuitive, since large values of $l$ correspond to long random walks,
that depend very little about the conditions that are enforced at the
endpoints of the path.

This observation has also very practical consequences in the
evaluation of the expectation of local operators that contain
derivatives. In the particular example of eq.~(\ref{eq:phi2-NL}), it
is easy to see that the second order coefficients in the small $a$
expansion will always come in combinations such as ${\bs
  D}_{4,\hat{1}}^{12,12}- {\bs C}_4^{12,12}$ or ${\bs
  D}_{4,\hat{3}}^{12,12}- {\bs C}_4^{12,12}$. The summands in these
differences have a milder behavior at large $l$, and instead of having
a quadratic infrared divergence, they have only a logarithmic
divergence
\begin{equation}
{\bs D}_{4,\hat{1}}^{12,12}- {\bs C}_4^{12,12}\sim{\bs
  D}_{4,\hat{3}}^{12,12}- {\bs C}_4^{12,12}\sim \log(l_{\rm max})\; ,
\end{equation}
as expected on dimensional grounds.

\subsection{Bilocal operators with a 2-hop separation}
In the discretization of $(D_\nu\phi)^*(D_\mu\phi)$, that appears for
instance in the e\-ner\-gy-momentum tensor, we also need to consider
operators such as
\begin{equation}
\phi_a^*(x+\hat{\nu})U_\nu^{ab\dagger}(x)U_\mu^{bc}(x)\phi_c(x+\hat{\mu})\; .
\end{equation}
When $\mu=\nu$, the two Wilson lines cancel and this operator is
simply the local operator $\phi^*\phi$ evaluated at the point
$x+\hat\mu$. The novel case is when $\mu\not=\nu$, for which the two
elementary fields are separated by two hops in distinct directions.

For the leading order coefficient ${\bs D}_{0,{\hat\nu}^{-1}\hat\mu}$, we can
use the result derived in the appendix \ref{app:NL-D0},
\begin{equation} 
{\bs D}_{0,{\hat\nu}^{-1}\hat\mu}
=
\int_0^\infty dt \; e^{-t}\; I_1^2(\tfrac{t}{d})I_{0}^{d-2}(\tfrac{t}{d})\; .
\label{eq:D0-int-munu}
\end{equation}

For the next-to-leading order coefficient ${\bs
  D}_{4,{\hat\nu}^{-1}\hat\mu}^{12;12}$, we need now to distinguish three
cases, depending on whether the directions $\mu,\nu$ coincide with the
directions $1,2$ of the plane on which the areas are projected. The
results for a single hop separation can be generalized into~:
\begin{itemize}
\item[{\bf i.}] $\mu,\nu\not= 1,2$~:
  \begin{equation}
    {\bs D}_{4,{\hat\nu}^{-1}\hat\mu}^{12;12}=\frac{1}{12 d^2}\int_0^\infty dt\;e^{-t}\; t^2\;
I_1^4(\tfrac{t}{d})\;
I_0^{d-4}(\tfrac{t}{d})\; ,
\label{eq:rw-xymunu}
  \end{equation}
\item[{\bf ii.}] $\nu=1, \mu\not= 1,2$~:
  \begin{equation}
    {\bs D}_{4,\hat{1}^{-1}\hat\mu}^{12;12}=\frac{1}{12 d^2}\int_0^\infty dt\;e^{-t}\; t^2\;
I_1^2(\tfrac{t}{d})\;
I_0^{d-2}(\tfrac{t}{d})\; ,
\label{eq:rw-xymux}
  \end{equation}
\item[{\bf iii.}] $\nu=1, \mu=2$~: by using eq.~(\ref{eq:area-nxy-0xy})
  and a Borel transformation, we can obtain the following expression~:
  \begin{equation}
    {\bs D}_{4,\hat{1}^{-1}\hat{2}}^{12;12}=\frac{1}{6}\int_0^\infty dt\;e^{-t}\;
I_0^{d-2}(\tfrac{t}{d})\;\left[\frac{t^2}{2d^2}I_0^2(\tfrac{t}{d})+I_1^2(\tfrac{t}{d})\right]\; .
\label{eq:rw-xyyx}
  \end{equation}

\end{itemize}

\section{Anisotropic lattice}
\label{sec:aniso}
\subsection{Worldline representation}
Let us now consider a lattice in which the spacings are not the same
in all directions: $a_1,a_2,\cdots, a_d$. The covariant derivative
squared now reads
\begin{eqnarray}
&&
-D^2 f(i_1,\cdots,i_d)
=
-2\Big[\sum_{r=1}^d\frac{1}{a_r^2}\Big]f(i_1,\cdots,i_d)
\nonumber\\
&&\quad
+\sum_{r=1}^d\left\{\frac{U_r(\cdots,i_r,\cdots)f(\cdots,i_r+1,\cdots)}{a_r^2}
\right.
\nonumber\\
&&\qquad\qquad
+\left.\frac{U_r^{-1}(\cdots,i_r-1,\cdots)f(\cdots,i_r-1,\cdots)}{a_r^2}\right\}\; .
\end{eqnarray}
Let us introduce a ``mean inverse squared lattice spacing'',
\begin{equation}
\frac{1}{{\bf a}^2}\equiv \frac{1}{d}\sum_{r=1}^d\frac{1}{a_r^2}\; ,
\end{equation}
and now we write
\begin{equation}
\frac{1}{D^2}
\equiv
\frac{{\bf a}^2}{2d}\sum_{n=0}^\infty
(1-{\bf a}^2 D^2/2d)^n\; .
\end{equation}
This has again the virtue of eliminating the ``stationary'' term in
the random walk, since we have
\begin{eqnarray}
&&
\Big(1-\frac{{\bf a}^2D^2}{2d}\Big) f(i_1,\cdots,i_d)
=
\frac{{\bf a}^2}{2d}\sum_{r=1}^d\left\{\frac{U_r(\cdots,i_r,\cdots)f(\cdots,i_r+1,\cdots)}{a_r^2}
\right.
\nonumber\\
&&\qquad\qquad\qquad\qquad\qquad
+\left.\frac{U_r^{-1}(\cdots,i_r-1,\cdots)f(\cdots,i_r-1,\cdots)}{a_r^2}\right\}\; .
\end{eqnarray}
The difference compared to the isotropic case is that the
probability to make a hop in the direction $r$ is modified by the
factor
\begin{equation}
h_r\equiv \frac{{\bf a}^2}{a_r^2}\; .
\end{equation}
This factor is trivially 1 if the spacings are all equal but differs
from 1 for an anisotropic lattice. The hopping probability in the
direction $r$ is inversely proportional to the squared lattice spacing
in this direction (the factor $h_r$ is almost equal to $d$ in the
direction that has the smallest lattice
spacing). Eq.~(\ref{eq:exp-phi2-2}) is thus generalized into
\begin{equation}
\big<\phi_a(0)\phi^*_a(0)\big>
=\frac{{\bf a}^2}{2d\prod_r a_r}\,\sum_{n=0}^\infty
\frac{1}{(2d)^{2n}}\;
\sum_{\gamma\in\Gamma_{2n}(0,0)} \Big[\prod_{\ell\in\gamma} h_\ell\Big]\;
{\rm tr}_{\rm adj}\,\big({\cal W}(\gamma)\big)\; ,
\label{eq:exp-phi2-3}
\end{equation}
where $\prod_{\ell\in\gamma} h_\ell$ denotes the product of all the $h$'s 
collected along the closed loop.

\subsection{Geometrical interpretation}
Note that the mean squared value of the absolute distance traveled in
the direction $r$ is given by
\begin{equation}
\big<\Delta \ell_r^2\big>=n_r\; a_r^2\; ,
\end{equation}
where $n_r$ is the number of hops in the direction $r$. Since $n_r$ is
proportional to $h_r$, the product $n_r a_r^2$ is independent of the
direction $r$, and we have
\begin{equation}
\big<\Delta \ell_1^2\big>
=
\big<\Delta \ell_2^2\big>
=\cdots=
\big<\Delta \ell_d^2\big>
\; ,
\end{equation}
regardless of the values of the lattice spacings. In other words, the
loops that enter in the formula (\ref{eq:exp-phi2-3}) have an absolute
geometrical shape which is isotropic (on average). On an anisotropic
lattice, this isotropic distribution of shapes is realized by making
more hops in the directions that have a smaller lattice spacing.

\subsection{Leading order coefficients ${\bs C}_0$ and ${\bs D}_{0,\hat{x}}$}
The coefficient that appears in the leading term of the continuum
limit is now modified into
\begin{equation}
{\bs C}_0(\{h_r\})\equiv \sum_{n=0}^\infty
\frac{1}{(2d)^{2n}}\sum_{\gamma\in\Gamma_{2n}(0,0)} \Big[\prod_{\ell\in\gamma} h_\ell\Big]\; .
\end{equation}
A somewhat more explicit expression can be obtained by partitioning
the $2n$ hops as $2n=2n_1+\cdots+2n_d$, where $2n_1$ is the number of
hops in the $x_1$ direction, $2n_2$ the number of hops in the $x_2$
direction, etc... This leads to
\begin{equation}
{\bs C}_0(\{h_r\})
=
\sum_{n=0}^\infty
\frac{(2n)!}{(2d)^{2n}}\sum_{n_1+\cdots+n_d=n}\frac{h_1^{2n_1}\cdots h_d^{2n_d}}{n1!^2\cdots n_d!^2}\; .
\end{equation}
One can easily generalize the derivation of eq.~(\ref{eq:C0-bessel})
in the anisotropic case. This leads to the following integral
representation for ${\bs C}_0$,
\begin{equation}
{\bs C}_0(\{h_r\})
=\int_0^\infty dt\;e^{-t}\;\prod_{r=1}^d I_0(\tfrac{h_r t}{d})\; .
\label{eq:C0-bessel-aniso}
\end{equation}
Similarly to eq.~(\ref{eq:C0-bessel-aniso}), it is possible to obtain
the following integral representation of the leading order coefficient
${\bs D}_{0,\hat{x}}$ that appears in the continuum expansion of
composite operators involving fields separated by one lattice spacing~:
\begin{equation}
{\bs D}_{0,\hat{1}}(\{h_r\})
=\int_0^\infty dt\;e^{-t}\;I_1(\tfrac{h_1 t}{d})\prod_{i\not=1} I_0(\tfrac{h_i t}{d})\; .
\label{eq:D0-bessel-aniso}
\end{equation}
Likewise, the anisotropic form of the formulas of the appendix
\ref{app:NL-D0} is completely straightforward.

In the example of the operator of eq.~(\ref{eq:phi2-NL}), the leading
term in the expansion in powers of the lattice coupling can be
expressed in terms of the coefficients ${\bs C}_0$ and ${\bs
  D}_{0,{\hat{x}}}$,
\begin{eqnarray}
\left<\phi^*_a(0)(D_\mu D_\mu\phi)_a(0)\right>
=
\frac{{\rm tr}_{\rm adj}\,(1)}{d}\frac{{\bf a}^2}{a_1\cdots a_{d}}
\sum_{i=1}^{d} \frac{{\bs D}_{0,{\hat{\imath}}}-{\bs C}_{0}}{a_i^2}+\cdots
\label{eq:phi2-NL-1}
\end{eqnarray}
Using the identities $I_1=I_0'$ and $(xI_1(x))'=xI_0(x)$, and
integrating by parts in eq.~(\ref{eq:D0-bessel-aniso}), we obtain
\begin{eqnarray}
\sum_{i=1}^d h_i\,{\bs D}_{0,\hat{\imath}}
=
d\,\big({\bs C}_0-1\big)\; .
\end{eqnarray}
This identity leads to
\begin{equation}
\sum_{i=1}^{d} \frac{{\bs D}_{0,{\hat{\imath}}}+1-{\bs C}_{0}}{a_i^2}=0\; .
\label{eq:Id-C0}
\end{equation}
When we use this property in eq.~(\ref{eq:phi2-NL-1}), it turns it into
\begin{equation}
\left<\phi^*_a(0)(D_\mu D_\mu\phi)_a(0)\right>
=
-\frac{{\rm tr}_{\rm adj}\,(1)}{a_1\cdots a_d}+\cdots\; ,
\label{eq:phi2-NL-2}
\end{equation}
which is consistent with the equation of motion satisfied by the
propagator~:
\begin{equation}
\left<\phi^*_a(0)(D_\mu D_\mu\phi)_a(0)\right>
=
\lim_{x\to 0}(D_\mu D_\mu)^{ab}_x\left<\phi^*_a(0)\phi_b(x)\right>
=
-{\rm tr}_{\rm adj}\,(1)\;\lim_{x\to 0}\delta(x)\; .
\end{equation}
A similar identity, eq.~(\ref{eq:Id-C4}), among the coefficients that
appear at the next order ensures that this property does not receive
corrections that depend on the background field (i.e. the dots in
eq.~(\ref{eq:phi2-NL-2}) are in fact zero).

\subsection{Coefficient ${\bs C}_4^{12;12}$}
In the isotropic case, an essential ingredient in the calculation of
the coefficient ${\bs C}_4$ was the combinatorial formula
(\ref{eq:A2-d2}), that gives the variance of the areas enclosed by
closed random walks in 2 dimensions. Now, we need to
generalize this formula to random walks weighted by the factors $h_i$
that depend on the direction of each hop. Note that in the isotropic
case ($h_i=1$), this quantity has a simple generating function,
\begin{equation}
\sum_{n=0}^\infty
\frac{X^{2n}}{(2n)!}
\left[\sum_{\gamma\in\Gamma_{2n}(0,0)} \left(A_{12}(\gamma)\right)^2\right]_{{\rm\ dim\ }2}
=
\frac{X^2}{3}\;I_1^2(2X)\; .
\label{eq:A2-gen-iso}
\end{equation}
The integral representation (\ref{eq:C0-bessel-aniso}) of ${\bs C}_0$
in the anisotropic case and its comparison with the corresponding
isotropic formula (\ref{eq:C4-bessel}) suggests the following
generalization of eq.~(\ref{eq:A2-gen-iso})~:
\begin{eqnarray}
&&\sum_{n=0}^\infty
\frac{X^{2n}}{(2n)!}
\left[\sum_{\gamma\in\Gamma_{2n}(0,0)} \Big[\prod_{l\in\gamma}h_l\Big]\;
\left(A_{12}(\gamma)\right)^2\right]_{{\rm\ dim\ }2}
\nonumber\\
&&\qquad\qquad\qquad\qquad=
\frac{h_1 h_2X^2}{3}\;I_1(2h_1X)\;I_1(2h_2X)\; .
\label{eq:A2-gen-aniso}
\end{eqnarray}
It is straightforward to see that this formula is equivalent to the
identity (\ref{eq:area-nxy-00}) listed in the appendix
\ref{app:area-aniso}.  In a sense, eq.~(\ref{eq:area-nxy-00}) can be
viewed as a ``fine grained'' version of eq.~(\ref{eq:A2-d2}), that
retains the information about the number of hops in each direction. We
present a proof of this identity in a separate paper~\cite{EpelbGW3}.

With the help of eq.~(\ref{eq:A2-gen-aniso}), it is immediate to
obtain the integral representation of the coefficient ${\bs
  C}_4^{12;12}$ in the anisotropic case,
\begin{equation}
{\bs C}_4^{12;12}(\{h_r\})=\frac{h_1 h_2}{12d^2}
\int_0^\infty dt\; e^{-t}\;t^2\;\Big[\prod_{i\not=1,2}I_0(\tfrac{h_it}{d})\Big]
\;I_1(\tfrac{h_1t}{d})\;I_1(\tfrac{h_2t}{d})\; .
\label{eq:C4-bessel-aniso}
\end{equation}
This coefficient will enter in the short distance expansion via the combination 
\begin{equation}
a_1^2 a_2^2\; {\bs C}_4^{12;12} F_a^{12}(0)F_a^{12}(0)\sim {\bf a}^4 F_a^{12}(0)F_a^{12}(0)\; ,
\end{equation}
where the unwritten factors are dimensionless numbers. Therefore, the
terms in $F^2$ are accompanied by the fourth power of the ``average
lattice spacing'' ${\bf a}$, with a numerical prefactor that depends
on the ratios $a_i/a_j$ of the lattice spacings in the various
directions. Thus, if we take the short distance limit while keeping
these ratios fixed, only the factor ${\bf a}^4$ decreases, while the
numerical prefactor stays constant.

\subsection{Coefficients ${\bs D}_{4,\hat{\mu}}^{12;12}$ and ${\bs D}_{4,\hat{\nu}^{-1}\hat{\mu}}^{12;12}$}
Likewise, eqs.~(\ref{eq:D4z-bessel}) and (\ref{eq:D4x-bessel}) can be
generalized into
\begin{equation}
{\bs D}_{4,\hat{3}}^{12;12}(\{h_r\})=
\frac{h_1 h_2}{12d^2}\int_0^\infty \!\!\!dt\; e^{-t}\; t^2\; 
\Big[\prod_{i\not=1,2,3}I_0(\tfrac{h_it}{d})\Big]\;
I_1(\tfrac{h_1t}{d})\,I_1(\tfrac{h_2t}{d})\,I_1(\tfrac{h_3t}{d})\; ,
\label{eq:D4z-bessel-aniso}
\end{equation}
and
\begin{equation}
{\bs D}_{4,\hat{1}}^{12;12}(\{h_r\})=
\frac{h_1 h_2}{12d^2}\int_0^\infty \!\!\!
dt\; e^{-t}\; t^2\; \Big[\prod_{i\not=2}I_0(\tfrac{h_it}{d})\Big]\;
I_1(\tfrac{h_2t}{d})\; .
\label{eq:D4x-bessel-aniso}
\end{equation}
Using these formulas, we can also prove the identity\footnote{A word
  of caution is necessary here: the l.h.s. and r.h.s. of this equation
  contain infrared divergences that should be properly regularized,
  for instance by the introduction of a mass (see the section
  \ref{sec:mass}).}
\begin{equation}
\sum_{i=1}^d h_i {\bs D}_{4,\hat{\imath}}^{12;12}=d\, {\bs C}_{4}^{12;12}\; ,
\label{eq:Id-C4}
\end{equation}
that generalizes eq.~(\ref{eq:Id-C0}) to the coefficients that appear
in the next order of the expansion.

It is also easy to generalize to the case an anisotropic lattice the
coefficients ${\bs D}_{4,\hat{\nu}^{-1}\hat{\mu}}^{12;12}$ with a
separation of two hops, given in eqs.(\ref{eq:rw-xymunu}),
(\ref{eq:rw-xymux}) and (\ref{eq:rw-xyyx}) in the isotropic case. One
obtains~:
\begin{itemize}
\item[{\bf i.}] $\mu,\nu\not= 1,2$~:
  \begin{equation}
    {\bs D}_{4,{\hat\nu}^{-1}\hat\mu}^{12;12}=
    \frac{h_1h_2}{12 d^2}\!\int_0^\infty \!\!\!\!dt\,e^{-t}\, t^2\,
I_1(\tfrac{h_1t}{d})\,I_1(\tfrac{h_2t}{d})\,I_1(\tfrac{h_\mu t}{d})\,I_1(\tfrac{h_\nu t}{d})
\!\!\!\!\prod_{i\not=1,2,\mu,\nu}\!\!\!\!I_0(\tfrac{h_it}{d})\; ,
\label{eq:rw-xymunu-aniso}
  \end{equation}
\item[{\bf ii.}] $\nu=1, \mu\not= 1,2$~:
  \begin{equation}
    {\bs D}_{4,\hat{1}^{-1}\hat\mu}^{12;12}=
    \frac{h_1h_2}{12 d^2}\int_0^\infty dt\;e^{-t}\; t^2\;
I_0(\tfrac{h_1 t}{d})\,I_1(\tfrac{h_2 t}{d})\,I_1(\tfrac{h_\mu t}{d})\,
\prod_{i\not=1,2,\mu}I_0(\tfrac{h_i t}{d})\; ,
\label{eq:rw-xymux-aniso}
  \end{equation}
\item[{\bf iii.}] $\nu=1, \mu=2$~:
  \begin{eqnarray}
    {\bs D}_{4,\hat{1}^{-1}\hat{2}}^{12;12}
   & =&
    \frac{1}{6}\int_0^\infty dt\;e^{-t}\;
\Big[\frac{h_1h_2t^2}{2d^2}I_0(\tfrac{h_1 t}{d})I_0(\tfrac{h_2 t}{d})
\nonumber\\ 
&&
\qquad\qquad\qquad+
I_1(\tfrac{h_1 t}{d})I_1(\tfrac{h_2 t}{d})\Big]
\prod_{i\not=1,2}I_0(\tfrac{h_i t}{d})\; .
\label{eq:rw-xyyx-aniso}
  \end{eqnarray}

\end{itemize}

\subsection{Limit of extreme anisotropy}
\label{sec:extreme-aniso}
An extreme case of anisotropy is to have one lattice spacing much
smaller than the others. This could correspond to the case where one
of the coordinates (e.g. the time) is treated as a continuous variable
while the other are discretized with a small but finite lattice
spacing. Let us therefore assume that
\begin{equation}
a_d\ll a_1,\cdots,a_{d-1}\to 0\; .
\end{equation}

\begin{figure}[htbp]
\begin{center}
\resizebox*{9cm}{!}{\includegraphics{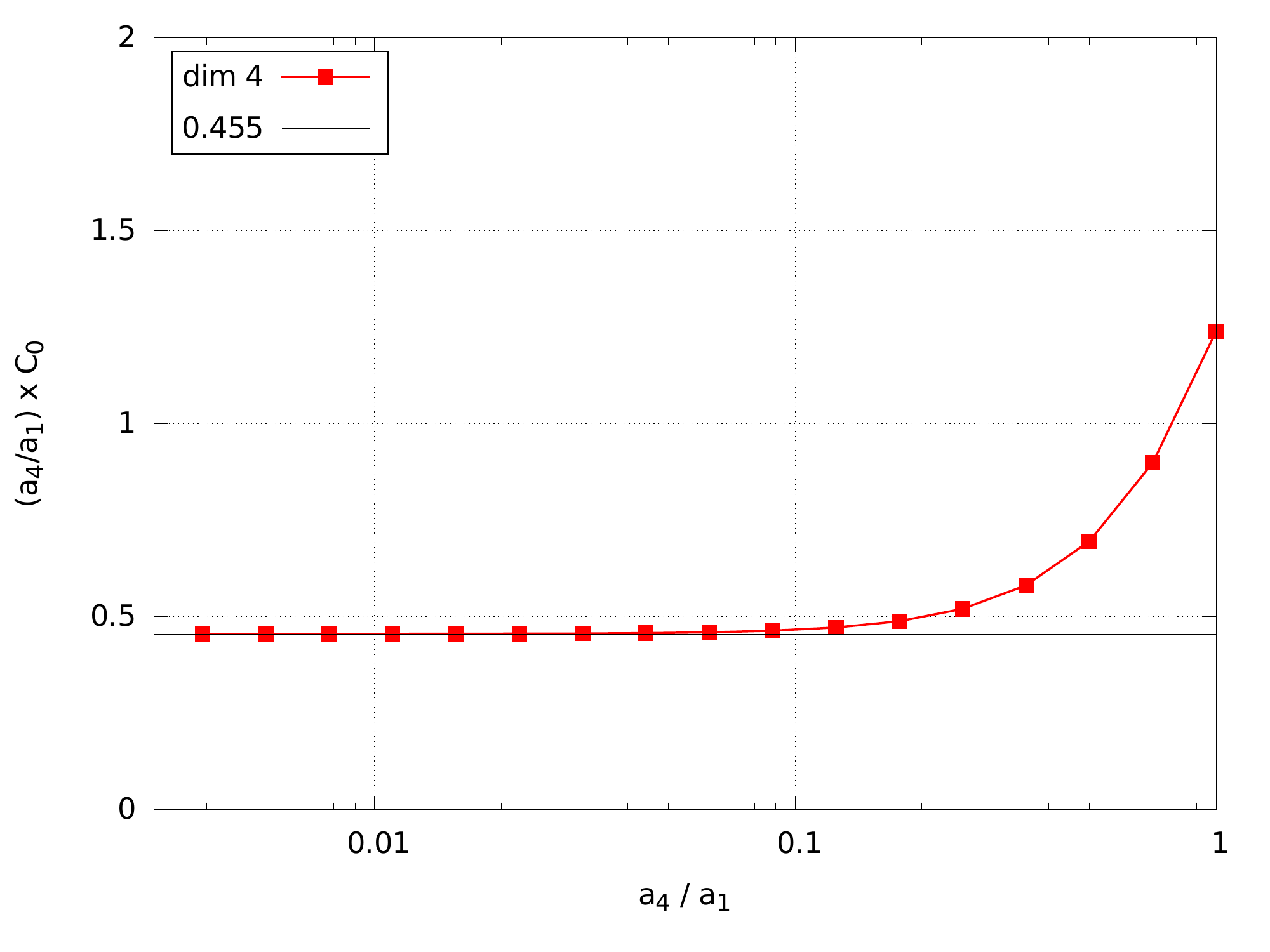}}
\end{center}
\caption{\label{fig:aniso-C0}Behavior of the leading coefficient ${\bs
    C}_0$ as a function of the ratio of lattice spacings $a_4/a_1$, in
  4 dimensions.}
\end{figure}

In order to get a sense of the behavior of the coefficients in this
limit, let us consider the coefficient ${\bs C}_0$, whose integral
representation (\ref{eq:C0-bessel-aniso}) can easily be studied
numerically. In this numerical study, we consider the dimension $d=4$,
and the remaining $3$ lattice spacings are assumed to be all equal. As
shown in the figure \ref{fig:aniso-C0}, we see that ${\bs C}_0$
diverges as $a_4^{-1}$ in this limit.
In this limit, we have
\begin{equation}
{\bf a}^2\approx 4a_4^2\quad,\qquad \frac{{\bf a}^2}{a_1a_2a_3a_4}\approx \frac{4a_4}{a_1a_2a_3}\; ,
\end{equation}
so that the leading term of eq.~(\ref{eq:exp-phi2-3}) in fact becomes
independent of the smallest lattice spacing. This result is consistent
with the result of the appendix \ref{app:cont-time}, where we use from
the start a continuous time variable.

It is in fact possible to understand this limit analytically, starting
from the integral representation of the coefficients that appear in
the small lattice spacing expansion. Generically, these coefficients
involve integrals of the form,
\begin{equation}
A_{n;\{\delta_i\}}\equiv\int_0^\infty dt\;e^{-t}\; t^n\;\prod_{i=1}^d I_{\delta_i}(\tfrac{h_i t}{d})\; ,
\end{equation}
where the exponent $n$ is 0 or 2 and the indices $\delta_i$ are 0 or
1, in all the examples we have encountered so far. In this limit, we
have
\begin{equation}
h_d=d\quad,\qquad h_i=\frac{d a_d^2}{a_i^2}\ll 1\quad(i<d)\; .
\end{equation}
Therefore, the argument of the last Bessel function is much larger
than the arguments of the first $d-1$ Bessel functions. Let us first
define a rescaled integration variable by
\begin{equation}
\tau = t\left(1-\frac{h_d}{d}\right)=t\,\frac{d_s a_d^2}{{\bf a}_s^2+d_s a_d^2}\; ,
\label{eq:tau}
\end{equation}
where we denote
\begin{equation}
d_s\equiv d-1\quad,\qquad \frac{1}{{\bf a}_s^2}\equiv\frac{1}{d_s}\sum_{i=1}^{d_s}\frac{1}{a_i^2}\; .
\end{equation}
Using $h_it/d={\bf a}_s^2\tau /(d_sa_i^2)$ and the asymptotic
expansion of the modified Bessel function,
\begin{equation}
I_{\delta}(z)\empile{=}\over{z\to+\infty}\frac{e^z}{\sqrt{2\pi}}\,\Big[\frac{1}{\sqrt{z}}+{\cal O}\big(z^{-3/2}\big)\Big]\; ,
\end{equation}
we arrive at
\begin{equation}
A_{n;\{\delta_i\}}\empile{=}\over{a_d\to 0}
\frac{1}{\sqrt{2\pi}}\left(\frac{{\bf a}_s^2}{d_s a_d^2}\right)^{n+\tfrac{1}{2}}
\int_0^\infty d\tau\;e^{-\tau}\; \tau^{n-1/2}\;\prod_{i=1}^{d_s} I_{\delta_i}(\tfrac{h_{is} \tau}{d_s})\; ,
\label{eq:int-ad-0}
\end{equation}
where we define $h_{is}\equiv {\bf a}_s^2/a_i^2$. Note that this
limiting value does not depend on the index $\delta_d$ of the Bessel
function associated to the temporal direction.

When $n=0$ (i.e. for the coefficients ${\bs C}_0$ and ${\bs D}_0$),
this integral behaves as $a_d^{-1}$, which is cancelled by the
behavior of the overall prefactor ${\bf a}^2/(a_1\cdots
a_d)$. Therefore, the leading order terms have a finite
limit\footnote{\label{foot:C0}If we take $a_1=\cdots=a_{d-1}\gg a_d$,
  then the leading term of $\big<\phi_a^*(0)\phi_a(0)\big>$ reads
\begin{equation*}
\big<\phi_a^*(0)\phi_a(0)\big>\empile{=}\over{a_d\ll a_{1\cdots d-1}\to 0}
\frac{{\rm tr}_{\rm adj}\,(1)}{2a_1^{d-2}}
\underbrace{\frac{1}{\sqrt{2\pi d_s}}\int_0^\infty\frac{d\tau}{\sqrt{\tau}}\;e^{-\tau}\;I_0^{d_s}(\tfrac{\tau}{d_s})}_{0.4553440518\cdots\ {\rm if\ }d_s=3}+\cdots\; ,
\end{equation*} in perfect agreement with the figure \ref{fig:aniso-C0}.} when $a_d\to 0$. In the next-to-leading
order coefficients ${\bs C}_4^{ij;ij}$ and ${\bs D}_4^{ij;ij}$, the
integral (\ref{eq:int-ad-0}) appear with the exponent $n=2$ and
therefore it behaves as $a_d^{-5}$. One of the powers of $a_d^{-1}$ is
cancelled by the prefactor ${\bf a}^2/(a_1\cdots a_d)$. Moreover, the
coefficients ${\bs C}_4^{ij;ij}$ and ${\bs D}_4^{ij;ij}$ also contain
a prefactor $h_i h_j$,
\begin{equation}
h_ih_j\empile{=}\over{a_d\to 0}
\frac{d^2a_d^4}{a_i^2 a_j^2}\; ,
\end{equation}
thereby canceling the remaining factor $a_d^{-4}$ from
eq.~(\ref{eq:int-ad-0}). Therefore, the next-to-leading order
coefficients are also finite in the limit $a_d\to 0$ (aside from
the possible infrared divergences discussed previously).

\subsection{Energy-momentum tensor}
\label{sec:tmunu}
Thanks to the results of the previous sections, we can write the short
distance expansion of the energy momentum tensor. In this section, we
present expressions for its diagonal components (its off-diagonal
components can be treated similarly, but have a slightly more
complicated structure). These diagonal components read~:
\begin{equation}
T^{ii}=(D_i\phi)^*(D_i\phi)-\sum_{j\not=i}(D_j\phi)^*(D_j\phi)\; .
\label{eq:Tii-1}
\end{equation}
Therefore, we need the expression of the expectation value of the
operator $(D_i\phi)^*(D_i\phi)$ (not summed over $i$).  The lattice
version of this operator is
\begin{equation}
(D_i\phi(0))^*(D_i\phi(0))=
\frac{
\phi^*(0)\phi(0)
\!+\!
\phi^*(\hat{\imath})\phi(\hat{\imath})
\!-\!
\phi^*(\hat{\imath})U_i^\dagger(0)\phi(0)
\!-\!
\phi^*(0)U_i(0)\phi(\hat{\imath})
}{a_i^2}\;,
\end{equation}
and the worldline representation of its 1-loop expectation value reads
\begin{eqnarray}
&&
\!\!\!\!\!\!\!\!
\big<(D_i\phi(0))^*(D_i\phi(0))\big>=
\frac{{\bf a}^2}{2d\prod_r a_r}\frac{1}{a_i^2}
\sum_{n=0}^\infty
\frac{1}{(2d)^n}\Bigg\{
\sum_{\gamma\in\Gamma_n(0,0)}\Big[\prod_{\ell\in\gamma}h_l\Big]
\,{\rm tr}_{\rm adj}\,\big({\cal W}(\gamma)\big)
\nonumber\\
&&
\qquad\qquad\qquad\qquad\qquad\qquad
+
\sum_{\gamma\in\Gamma_n(\hat\imath,\hat\imath)}\Big[\prod_{\ell\in\gamma}h_l\Big]
\,{\rm tr}_{\rm adj}\,\big({\cal W}(\gamma)\big)
\nonumber\\
&&
\qquad\qquad\qquad\qquad\qquad\qquad
-
\sum_{\gamma\in\Gamma_n(\hat\imath,0)}\Big[\prod_{\ell\in\gamma}h_l\Big]
\,{\rm tr}_{\rm adj}\,\big(U_i^\dagger(0){\cal W}(\gamma)\big)
\nonumber\\
&&
\qquad\qquad\qquad\qquad\qquad\qquad
-
\sum_{\gamma\in\Gamma_n(0,\hat\imath)}\Big[\prod_{\ell\in\gamma}h_l\Big]
\,{\rm tr}_{\rm adj}\,\big(U_i(0){\cal W}(\gamma)\big)
\Bigg\}\; .
\end{eqnarray}
Using results derived earlier in this paper, we can write the
following short distance expansion for this quantity~:
\begin{eqnarray}
&&
\big<(D_i\phi(0))^*(D_i\phi(0))\big>=
\frac{{\bf a}^2}{2d\prod_r a_r}\,\frac{2}{a_i^2}\Bigg\{
{\rm tr}_{\rm adj}\,(1)\;\big({\bs C}_0-{\bs D}_{0,\hat\imath}\big)
\nonumber\\
&&
\qquad\qquad
-\frac{g^2}{4}\sum_{\mu<\nu}a_\mu^2 a_\nu^2 \big(F_a^{\mu\nu}(0)\big)^2\;
\big({\bs C}_4^{\mu\nu;\mu\nu}-{\bs D}_{4,\hat\imath}^{\mu\nu;\mu\nu}\big)
+
\cdots
\Bigg\}\; .
\end{eqnarray}
Using eq.~(\ref{eq:Tii-1}), as well as eqs.~(\ref{eq:Id-C0}) and
(\ref{eq:Id-C4}), we obtain the following expansion for the diagonal
components of the energy-momentum tensor,
\begin{eqnarray}
&&
T^{ii}=
\frac{{\bf a}^2}{d\prod_r a_r}\,\Bigg\{
{\rm tr}_{\rm adj}\,(1)\;
\Big(\frac{2}{a_i^2}\big({\bs C}_0-{\bs D}_{0,\hat\imath}\big)-\frac{d}{{\bf a}^2}\Big)
\nonumber\\
&&
\qquad\qquad
-\frac{g^2}{2a_i^2}\sum_{\ontop{k<l}{k,l\not=i}}a_k^2 a_l^2 \big(F_a^{kl}(0)\big)^2\;
\big({\bs C}_4^{kl;kl}-{\bs D}_{4,\hat\imath}^{kl;kl}\big)
\nonumber\\
&&
\qquad\qquad
-\frac{g^2}{2a_i^2}\sum_{k\not=i}a_i^2 a_k^2 \big(F_a^{ik}(0)\big)^2\;
\big({\bs C}_4^{ik;ik}-{\bs D}_{4,\hat\imath}^{ik;ik}\big)
+
\cdots
\Bigg\}\; ,
\label{eq:Tii-2}
\end{eqnarray}
where we have explicitly separated the terms of order $a^4$ depending
on whether one of the indices carried by the field strength is $i$ or
not.  Integral expressions in terms of modified Bessel functions for
all the coefficients that appear in this formula, for arbitrary
lattice spacings, can be found in the previous sections of this paper.
By using eqs.~(\ref{eq:Id-C0}) and (\ref{eq:Id-C4}), we can see that
\begin{equation}
\sum_{i=1}^d T^{ii}=(2-d)\frac{{\rm tr}_{\rm adj}\,(1)}{\prod_r a_r}+\cdots\; ,
\end{equation}
where the dots are UV finite terms in four dimensions.

Eq.~(\ref{eq:Tii-2}) provides the final answer for generic lattice
spacings. If in addition we assume that the temporal spacing $a_d$ is
much smaller than the others, we can use
\begin{eqnarray}
&&
\frac{2}{a_i^2}\big({\bs C}_0-{\bs D}_{0,\hat\imath}\big)-\frac{d}{{\bf a}^2}
\empile{=}\over{a_d\to 0}
-\frac{1}{a_d^2}\nonumber\\
&&\qquad
+
\left(\frac{2\,{\bf a}_s^2}{\pi d_s a_d^2}\right)^{1/2}\frac{1}{a_i^2}\int_0^\infty\frac{d\tau}{\sqrt{\tau}}\,e^{-\tau}\,\Big[I_0(\tfrac{h_{is}\tau}{d_s})-I_1(\tfrac{h_{is}\tau}{d_s})\Big]
\prod_{j\not=i,d}I_0(\tfrac{h_{js}\tau}{d_s})\nonumber\\
&&
\frac{2}{a_d^2}\big({\bs C}_0-{\bs D}_{0,\hat{d}}\big)-\frac{d}{{\bf a}^2}
\empile{=}\over{a_d\to 0}\frac{1}{a_d^2}
\; ,
\label{eq:C0-D0}
\end{eqnarray}
respectively for $i\not=d$ and $i=d$.  For the next-to-leading order
coefficients, we can also use eq.~(\ref{eq:int-ad-0}) to obtain their
limiting value if $i\not=d$. If $i=d$, it is necessary to go one order
further in the asymptotic expansion of the Bessel function of argument
$h_dt/d$. In this case, we can use the following limit
\begin{eqnarray}
&&
\frac{1}{a_d^2}
\int_0^\infty dt\;e^{-t}\; t^2\;\Big[I_0(\tfrac{h_d t}{d})-I_1(\tfrac{h_d t}{d})\Big]\prod_{i=1}^{d-1} I_{\delta_i}(\tfrac{h_i t}{d})
\nonumber\\
&&
\quad\empile{=}\over{a_d\to 0}
\frac{1}{\sqrt{8\pi}}\frac{d_s}{{\bf a}_s^2}\left(\frac{{\bf a}_s^2}{d_s a_d^2}\right)^{5/2}
\int_0^\infty d\tau\;e^{-\tau}\; \tau^{1/2}\;\prod_{i=1}^{d_s} I_{\delta_i}(\tfrac{h_{is} \tau}{d_s})\; .
\label{eq:int-bd-0}
\end{eqnarray}
The terms $\pm a_d^{-2}$ in eq.~(\ref{eq:C0-D0}) lead to
contributions pro\-por\-tio\-nal to $(a_1\cdots a_d)^{-1}$. It is easy to
check that all the other contributions to $T^{ii}$ become independent
of $a_d$ in the limit $a_d\to 0$.

\section{Conclusions}
\label{sec:conclusions}
In this paper, we have applied a discrete version of the worldline
formalism in order to obtain expressions for 1-loop expectation values
in a lattice scalar field theory, in the presence of a non-Abelian
gauge background. In this framework, 2-point correlators are expressed
as sums over all the random walks that connect their endpoints (or
closed loops in the specific case of local operators).  This
representation renders the ultraviolet and infrared behaviors of these
expectation values very intuitive: ultraviolet divergences are encoded
in the contribution of very short random walks, while the infrared
behavior arises from the statistics of long random walks that explore
large regions of space-time.

Moreover, it is straightforward to take the limit of small lattice
spacing in these worldline expressions. This gives an expansion in
powers of the background field strength, whose coefficients are sums
of all the closed random walks on the lattice, thereby reducing the
calculation of the coefficients to a combinatorial problem. The
leading coefficient is merely counting these random walks, and is
therefore very easy to obtain. The next-to-leading order coefficient
is related to the variance of the areas (projected on a plane)
enclosed by these random walks. The relevant combinatorial formulas
can be found in ref.~\cite{MingoN1}.

When one considers a lattice with anisotropic lattice spacings, these
random walks are further weighted by factors that count the number of
hops in each direction. In order to obtain simple expressions for the
coefficients of the expansion in this case, we had to conjecture some
generalizations of the formulas of ref.~\cite{MingoN1}. The proof of
these formulas is given in a separate paper~\cite{EpelbGW3}.

Using a Borel transformation, all the coefficients that appear in the
short distance expansion of these correlators can be rewritten as
1-dimensional integrals. These can be easily evaluated numerically,
and they are also quite convenient in order to study analytically the
limit where one lattice spacing becomes much smaller than the others.

This work can be extended in several directions. One of them is to
depart from a scalar field theory, in order to study quantities where
a spin $1/2$ fermion or a spin $1$ gauge boson circulates in the
loop. In the continuum theory, these extensions are well
known. Another --considerably more difficult-- extension would be to
use the worldline formalism in order to go beyond one loop. Also, at
the moment, it is unclear whether one could modify this formalism to
handle a Minkowskian time (see the discussion in the appendix
\ref{app:cont-time}).

\section*{Acknowledgements}
This work is supported by the Agence Nationale de la Recherche project
11-BS04-015-01.

\appendix

\section{Closed random walks on a periodic lattice}
\label{app:Pn}
In the main body of this paper, all the combinatorial formulas that we
have derived count random walks on an infinite cubic lattice. Since
the lattice is infinite, there is no need to specify boundary
conditions. This is the appropriate setting when one considers the
limit $a\to 0$ at fixed physical volume: the constant volume is
achieved by letting the number of lattice points go to infinity (as
$a^{-1}$) in each direction. In this appendix, we consider a different
limit, where we keep fixed the number of lattice points as we take the
limit $a\to 0$. We now assume periodic boundary conditions.

We have introduced in eq.~(\ref{eq:Iter}) a sequence of functions
$P_n$ defined on the lattice. If the first of these functions is
\begin{equation}
P_0(x)=\delta_{x,0}\; ,
\end{equation}
then $P_n(x)$ is the probability that a random walk starting at the
point $0$ reaches the point $x$ after $n$ steps\footnote{Using
eq.~(\ref{eq:Iter}) and ``integrating by parts'', it is easy to check
that the quantity
\begin{equation*}
\sum_{y\in{\rm lattice}}P_n(y)
\end{equation*} 
is conserved as $n$ increases.}. With this choice of initial
condition, $P_n(0)$ is the probability that the random walk makes a
closed loop of base point $0$ in $n$ steps.

The first step is to find the spectrum of the linear operator that
maps $P_n$ to $P_{n+1}$. By using eq.~(\ref{eq:Iter-1}), one finds
that the eigenfunctions and eigenvalues of this operator are the plane
waves
\begin{equation}
\phi_{{\vec k},d}({x})\equiv e^{2i\pi\frac{\{\vec k\cdot\vec x\}_d}{N}}\quad,\qquad
\{\vec k\cdot\vec x\}_d\equiv k_1 x_1+\cdots+k_d x_d\; ,
\end{equation}
and that the associated eigenvalue is
\begin{equation}
\Omega_{{\vec k},d}\equiv \frac{1}{d}\sum_{r=1}^d\cos\left(\frac{2\pi k_r}{N}\right)\; ,
\end{equation}
where $N$ is the number of lattice spacings in each direction (for
simplicity, we assume the same number of spacings in all the
directions). The initial condition of the iteration can be written as
\begin{equation}
P_0({x})=\frac{1}{N^d}\sum_{\{\vec k\}_d}\phi_{{\vec k},d}({x})\; ,
\end{equation}
where $\{\vec k\}_d$ denotes all the $d$-uplets of integers in the
range $[0,N-1]$. After $n$ iterations this distribution has become
\begin{equation}
P_n({x})=\frac{1}{N^d}\sum_{\{\vec k\}_d}\Omega_{{\vec k},d}^n\;\phi_{{\vec k},d}({x})\; .
\end{equation}
In order to obtain the probability for a random loop of length $n$ to
be closed, we simply evaluate this at $x=0$,
\begin{equation}
P_n({0})=\frac{1}{N^d}\sum_{\{\vec k\}_d}\Omega_{{\vec k},d}^n\; .
\label{eq:proba-closed}
\end{equation}
The sum over $n$ of the probability of closed random walks of length
$n$ therefore reads
\begin{equation}
\sum_{n\ge 0}P_n(0)=\frac{1}{N^d}\sum_{\{\vec k\}_d}\frac{1}{1-\Omega_{{\vec k},d}}
\; ,
\end{equation}
where we recognize the lattice expression for a 1-loop tadpole.

Let us now consider in more detail eq.~(\ref{eq:proba-closed}). We can make the right hand side more explicit by writing
\begin{eqnarray}
\frac{1}{N^d}\!\sum_{\{\vec k\}_d}\!\Omega_{{\vec k},d}^n
&=&
\frac{1}{N^d(2d)^n}\!\sum_{\{\vec k\}_d}\!
\Big(
e^{2i\pi\frac{k_1}{N}}+e^{-2i\pi\frac{k_1}{N}}
+\cdots+
e^{2i\pi\frac{k_d}{N}}+e^{-2i\pi\frac{k_d}{N}}
\Big)^n\nonumber\\
&=&\frac{1}{N^d(2d)^n}\!\sum_{\{\vec k\}_d}\!\sum_{\ontop{n_1+p_1+\cdots}{\ \ +n_d+p_d=n}}
\frac{n!}{n_1!p_1!\cdots n_d! p_d!}\nonumber\\
&&\qquad\qquad\qquad\qquad\times\;
e^{2i\pi\frac{k_1(n_1-d_1)}{N}}\cdots e^{2i\pi\frac{k_d(n_d-d_d)}{N}}\nonumber\\
&=&
\frac{1}{(2d)^n}\sum_{\ontop{n_1+p_1+\cdots}{\ \ +n_d+p_d=n}}
\frac{n!}{n_1!p_1!\cdots n_d! p_d!}
\delta_{n_1-p_1,0[N]}\cdots\delta_{n_d-p_d,0[N]}\; ,\nonumber\\
&&
\label{eq:000}
\end{eqnarray}
where the symbol $\delta_{p,0[N]}$ is a Kronecker symbol that defines
an equality modulo $N$~:
\begin{equation}
\delta_{p,0[N]}
\equiv
\sum_{i\in{\mathbbm Z}} \delta_{p,iN}\; .
\end{equation}
The essential difference between an infinite lattice and a finite
lattice with periodic boundary conditions comes from here.

If we take the limit $N\to\infty$, then we have
\begin{equation}
\delta_{p,0[N]}
\empile{=}\over{N\to\infty}
\delta_{p,0}\; ,
\end{equation}
and eq.~(\ref{eq:000}) is equivalent to
\begin{equation}
\frac{1}{N^d}\!\sum_{\{\vec k\}_d}\!\Omega_{{\vec k},d}^n
=\left\{
\begin{aligned}
&\frac{1}{(2d)^{2m}}\sum_{n_1+\cdots+n_d=m}
\frac{(2m)!}{n_1!^2\cdots n_d!^2}&&&(n=2m\mbox{\ even})\\
&0&&&(n \mbox{\ odd})\\
\end{aligned}
\right.\; ,
\end{equation}
which is nothing but the combinatorial expression for the coefficient
${\bs C}_0$ on an infinite cubic lattice.

In order to illustrate explicitly the difference on a finite periodic
lattice, let us consider a 1-dimensional lattice of $N$
sites. Eq.~(\ref{eq:000}) becomes
\begin{equation}
\frac{1}{N}\!\sum_{\{\vec k\}_1}\!\Omega_{{\vec k},1}^n
=
\frac{1}{2^n}\sum_{p=0}^n
\frac{n!}{p!(n-p)!}
\delta_{2p,n[N]}\; .
\end{equation}
For $N$ even, this forces the random walk to have an even length
$n=2m$, and we get
\begin{eqnarray}
\frac{1}{N}\!\sum_{\{\vec k\}_1}\!\Omega_{{\vec k},1}^{2m}
&\empile{=}\over{N{\rm\ even}}&
\frac{1}{2^{2m}}
\Big[
\frac{(2m)!}{m!^2}
+
2\frac{(2m)!}{(m+\frac{N}{2})!(m-\frac{N}{2})!}
\nonumber\\
&&\qquad\qquad
+
2\frac{(2m)!}{(m+\frac{2N}{2})!(m-\frac{2N}{2})!}
+\cdots
\Big]\; ,
\end{eqnarray}
where the sum in the right hand side stops when the argument of the
second factorial in the denominator becomes negative. The first term
of this formula is identical to the result on an infinite lattice. The
additional terms corresponds to paths that wrap around the periodic
lattice, with winding number $\pm 1$ for the second term, $\pm 2$ for
the third term, etc...  These terms can be viewed as finite size
corrections, since they explicitly depend on $N$. When the lattice
size $N$ is odd, the situation is even more complicated because random
walks of odd length are also permitted. For $n=2m$, we get
\begin{eqnarray}
\frac{1}{N}\!\sum_{\{\vec k\}_1}\!\Omega_{{\vec k},1}^{2m}
&\empile{=}\over{N{\rm\ odd}}&
\frac{1}{2^{2m}}
\Big[
\frac{(2m)!}{m!^2}
+
2\frac{(2m)!}{(m+{N})!(m-{N})!}
\nonumber\\
&&\qquad\qquad
+
2\frac{(2m)!}{(m+{2N})!(m-{2N})!}
+\cdots
\Big]\; ,
\end{eqnarray}
and for odd lengths $n=2m+1$ we have
\begin{eqnarray}
\frac{1}{N}\!\sum_{\{\vec k\}_1}\!\Omega_{{\vec k},1}^{2m+1}
&\empile{=}\over{N{\rm\ odd}}&
\frac{1}{2^{2m}}
\Big[
2\frac{(2m+1)!}{(m+\frac{1+N}{2})!(m+\frac{1-N}{2})!}
\nonumber\\
&&\qquad\qquad
+
2\frac{(2m+1)!}{(m+\frac{1+2N}{2})!(m+\frac{1-2N}{2})!}
+\cdots
\Big]\; .
\end{eqnarray}
The situation in $d$ dimensions follows the same pattern, but explicit
formulas become quite cumbersome. In particular, since the random walk
takes place on a $d$-dimensional torus instead of a ring, the
``winding number'' is now a $d$-uplet $(w_1,\cdots,w_d)\in{\mathbbm
  Z}^d$. For a given length $n$ of random walk, the term of winding
number $(0,\cdots,0)$ is identical to the result on an infinite
lattice, but in addition one must sum over all the possible windings
allowed for a given length $n$ of the random walk.

\section{Coefficient ${\bs D}_{0,\gamma_{x0}}$ for arbitrary separations}
\label{app:NL-D0}
In this appendix, we generalize the
eqs.~(\ref{eq:D0-1})--(\ref{eq:D0-4}) to the case where the points $0$
and $x$ are separated by more than one hop.  For definiteness, let us
assume that the point $x$ is
\begin{equation}
x\equiv x_1 \hat{1}+\cdots + x_d \hat{d}\; .
\end{equation}
The minimal length of a path connecting 0 to $x$ is:
\begin{equation}
\Delta\equiv x_1+\cdots+x_d\; .
\end{equation}
Firstly, let us notice that the parity of the length of the paths
$\gamma_{x0}$ that connect the two points depends solely on the
endpoints 0 and $x$. This parity is that of the number $\Delta$. One
can first obtain a combinatorial expression for ${\bs
  D}_{0,\gamma_{x0}}$,
\begin{equation}
{\bs D}_{0,\gamma_{x0}}=\sum_{n=\Delta}^\infty \frac{1}{(2d)^n}
\sum_{\Delta+2(n_1+\cdots+n_d)=n}
\frac{n!}{n_1!(n1+x_1)!\cdots n_d!(n_d+x_d)!}\; .
\end{equation}
By a Borel transformation, this can be turned into

\begin{equation}
{\bs D}_{0,\gamma_{x0}}=
\int_0^\infty dt \;e^{-t}\;B_d(\tfrac{t}{2d})\; ,
\end{equation}
with
\begin{equation}
B_d(x)\equiv \prod_{i=1}^d\left[
\sum_{p=0}^\infty \frac{x^{2p+x_i}}{p!(p+x_i)!}
\right]=\prod_{i=1}^d I_{x_i}(2x)\; .
\end{equation}
This leads to the following integral representation of ${\bs
  D}_{0,\gamma_{x0}}$,
\begin{equation} 
{\bs D}_{0,\gamma_{x0}}
=
\int_0^\infty dt \; e^{-t}\; \prod_{i=1}^d I_{x_i}(\tfrac{t}{d})\; .
\label{eq:D0-int}
\end{equation}
In 3 dimensions, this leads to a closed expression (see
ref.~\cite{GradsR1}--\S6.612.6)
\begin{equation}
{\bs D}_{0,\gamma_{x0}}
\empile{=}\over{d=3}
3\left(r_1 {\mathfrak g}+\frac{r_2}{\pi^2 {\mathfrak g}}+r_3\right)\; ,
\end{equation}
with
\begin{equation}
{\mathfrak g}\equiv \frac{\sqrt{3}-1}{96\pi^3}\,\Gamma^2\big(\tfrac{1}{24}\big)\,
\Gamma^2\big(\tfrac{11}{24}\big)\; ,
\end{equation}
and where the coefficients $r_{1,2,3}$ are listed in the table
\ref{tab:r} as a function of $x_{1,2,3}$ for small separations.
\begin{table}
\begin{center}
\begin{tabular}{c|ccc}
$x_1x_2x_3$& $r_1$& $r_2$& $r_3$\\
\hline
000&1&0&0\\
100&1&0&-1/3\\
110&5/12&-1/2&0\\
200&10/3&-2&-2\\
111&-1/8&3/4&0\\
210&3/8&-9/4&1/3\\
300&35/2&21&-13\\
\end{tabular}
\end{center}
\caption{\label{tab:r}Values of the coefficients $r_{1,2,3}$ for separations up to three hops (from \cite{GradsR1}--\S6.612.6). These values are invariant under permutations of $x_{1,2,3}$. For larger separations, see \cite{GradsR1}--\S6.612.6.}
\end{table}
In 4 dimensions, the integral of eq.~(\ref{eq:D0-int}) must be
evaluated numerically. Some values for small separations are listed in
the table \ref{tab:D0-4d}.

\begin{table}
\begin{center}
\begin{tabular}{c|ccc}
$x_1x_2x_3x_4$& ${\bs D}_{0,x_{1,2,3,4}}\quad(d=4)$\\
\hline
0000&1.239467122\\
1000&0.2394671218\\
1100&0.1017176302\\
2000&0.06596407193\\
1110&0.06187238110\\
2100&0.04365863661\\
3000&0.02629363394\\
\end{tabular}
\end{center}
\caption{\label{tab:D0-4d}Values of the leading order coefficient ${\bs
    D}_{0,\gamma_{x0}}$ in 4 dimensions, for separations
  up to three hops. These values are invariant under permutations of
  $x_{1,2,3,4}$.}
\end{table}

\section{Moments of the distribution of areas in $d=2$}
\label{app:area-aniso}
\subsection{Variance of the areas of closed loops with fixed sections}
In the case of bilocal operators, the next-to-leading order
coefficients ${\bs D}_{4,\gamma_{x0}}^{12;12}$ require that we
evaluate the variance of the areas enclosed by the paths
$\gamma\otimes\gamma_{x0}$, where $\gamma$ is a 2-dimensional random
walk from the point $0$ to the point $x$. In other words,
$\gamma\otimes\gamma_{x0}$ is a closed path, but only the section
$\gamma$ is random, while the section $\gamma_{x0}$ is held fixed.

\begin{figure}[htbp]
\begin{center}
\hfil
\resizebox*{5cm}{!}{\includegraphics{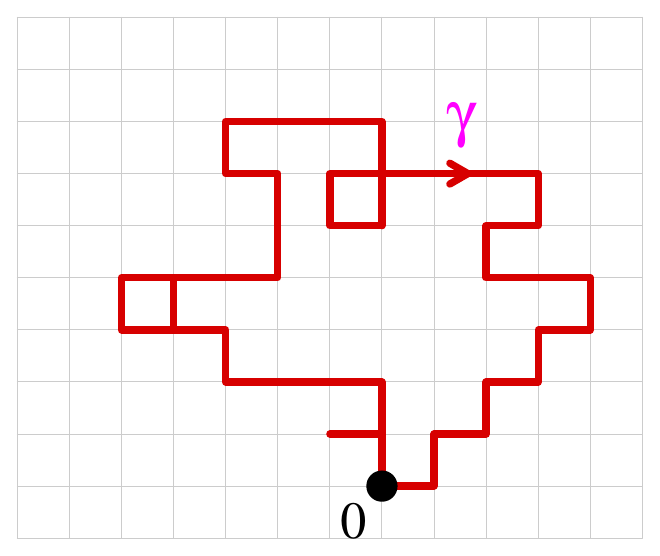}}
\hfill
\resizebox*{5cm}{!}{\includegraphics{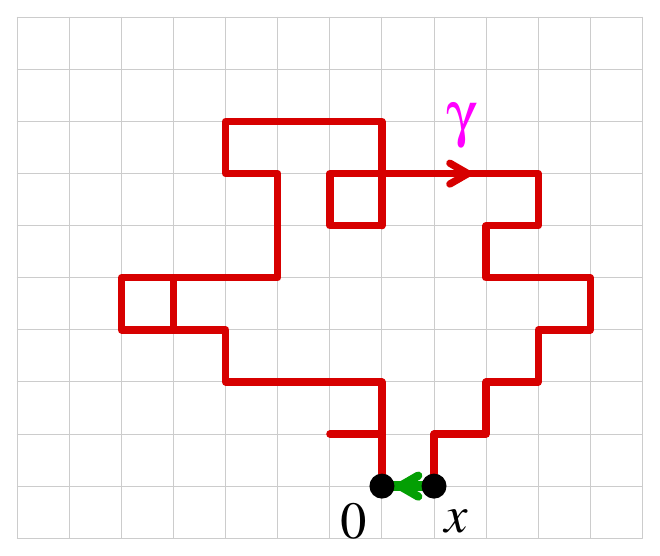}}
\hfill
\end{center}
\caption{\label{fig:rw-0-1}Left: closed random walk from $0$ to
  $0$. Right: closed random walk with a fixed section of length 1
  between $0$ and $x$ (shown in green).}
\end{figure}

\begin{figure}[htbp]
\begin{center}
\hfil
\resizebox*{5cm}{!}{\includegraphics{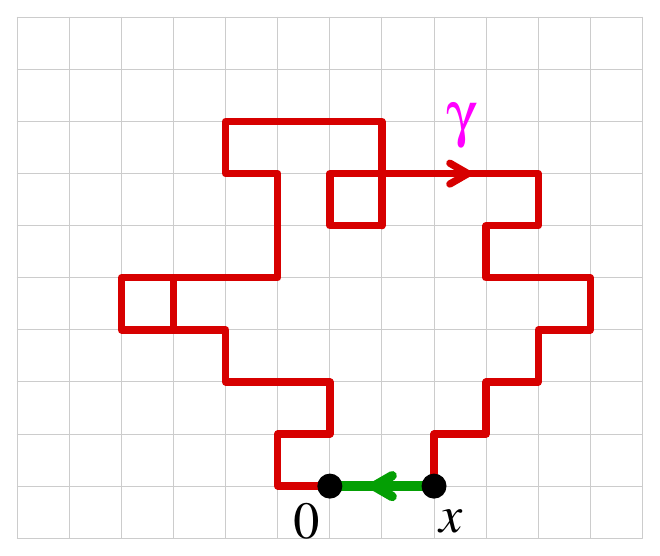}}
\hfill
\resizebox*{5cm}{!}{\includegraphics{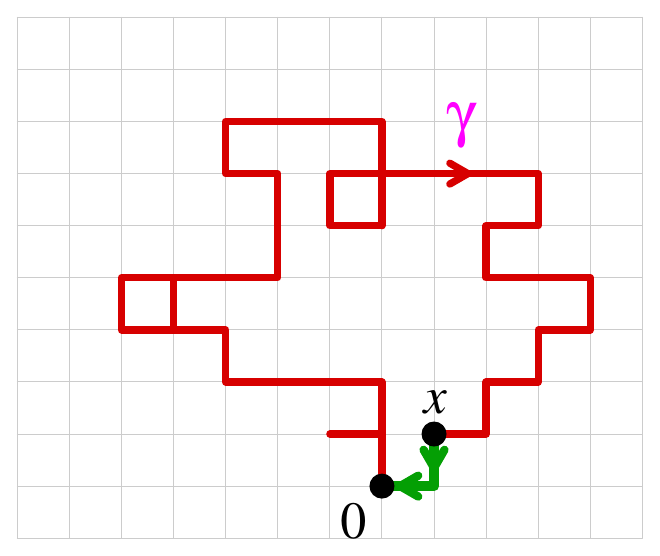}}
\hfill
\end{center}
\caption{\label{fig:rw-2}Closed random walks with a fixed section of
  length 2 (shown in green). Left: $x=2\,\hat{1}$. Right:
  $x=\hat{1}+\hat{2}$.}
\end{figure}

Having in mind the case of an anisotropic lattice, it is also useful
to have formulas that keep track of the number of hops in the $x_1$
and $x_2$ directions separately. All the formulas listed in this
appendix have been checked numerically by an exhaustive sum over all
random walks of length $2n\le 20$. In ref.~\cite{EpelbGW3}, we present
a proof of all the formulas of the next subsection, where the two
endpoints are identical. The same method can be used to study the case
where the endpoints are distinct, although we have not yet done so.

In the following, we give formulas for the four cases illustrated in
the figures \ref{fig:rw-0-1} and \ref{fig:rw-2}, which covers all the
situations where $0$ and $x$ are separated by two hops at most.  The
simplest situation is when the point $x=0$ and the path $\gamma_{x0}$
is the null path (figure \ref{fig:rw-0-1}, left). In this case, we
have\footnote{One can check that\begin{equation*} \sum_{n_1=0}^n
    \frac{(2n)!}{n_1!^2 (n-n_1)!^2}\;\frac{n_1 (n-n_1)}{3} =
    \frac{(2n)!^2}{n!^4}\;\frac{n^2(n-1)}{6(2n-1)}\; ,
  \end{equation*}which is nothing but the isotropic result for the variance of the areas in the set of all the closed random walks of lengths $2n$ (see \cite{MingoN1}, eqs.~(1.4)--(1.5)).}
\begin{equation}
\sum_{\gamma\in {\bs\Gamma}_{n_1,n_2}(0,0)}\big({\rm Area}\,(\gamma)\big)^2
=
\frac{(2(n_1+n_2))!}{n_1!^2 n_2!^2}\;\frac{n_1 n_2}{3}\; .
\label{eq:area-nxy-00}
\end{equation}
(We recall that ${\bs\Gamma}_{n_1,n_2}(0,x)$ is the set of all the
paths that connect the point 0 to the point $x$ and have $n_1$ hops in
the direction $+x_1$ and $n_2$ hops in the direction $+x_2$.)  

Next, let us consider the case where $x=\hat{x}_1$ and
$\gamma_{x0}=\hat{x}_1^{-1}$ (i.e. a single hop in the $-x_1$
direction -- see the figure \ref{fig:rw-0-1}, right). This case leads
to
\begin{equation}
\sum_{\gamma\in {\bs\Gamma}_{n_1,n_2}(0,\hat{1})}\big({\rm Area}\,(\gamma\otimes \hat{1}^{-1})\big)^2
=
\frac{(2(n_1+n_2)-1)!}{n_1!(n_1-1)! n_2!^2}\;\frac{n_1 n_2}{3}\; .
\label{eq:area-nxy-0x}
\end{equation}
For separations of two hops, there are two possibilities, shown in the
figure \ref{fig:rw-2}. When the point $x$ is located at $x=2\,\hat{1}$
and the path connecting to 0 is made of two horizontal hops (figure
\ref{fig:rw-2}, left), we have
\begin{equation}
\sum_{\gamma\in {\bs\Gamma}_{n_1,n_2}(0,2\,\hat{1})}\big({\rm Area}\,(\gamma\otimes \hat{1}^{-2})\big)^2
=
\frac{(2(n_1+n_2-1))!}{n_1!(n_1-2)! n_2!^2}\;\frac{(n_1+1) n_2}{3}\; .
\label{eq:area-nxy-02x}
\end{equation}
When the point $x$ is at $x=\hat{1}+\hat{2}$, and is connected to
the point 0 by the path $\gamma_{x0}=\hat{2}^{-1}\hat{1}^{-1}$,
we obtain
\begin{equation}
\sum_{\gamma\in {\bs\Gamma}_{n_1,n_2}(0,\hat{1}+\hat{2})}\big({\rm Area}\,(\gamma\otimes \hat{2}^{-1}\hat{1}^{-1})\big)^2
=
\frac{(2(n_1+n_2-1))!}{n_1!(n_1-1)! n_2!(n_2-1)!}\;\Big(\frac{n_1 n_2}{3}+\frac{1}{6}\Big)\; .
\label{eq:area-nxy-0xy}
\end{equation}

\subsection{Higher moments (for $x=0$ and $\gamma_{x0}={\bs 1}$)}
In the case of closed paths $\gamma$, it is also possible to obtain by
empirical observation a simple formula for the moment of order 4 of
the area,
\begin{equation}
\sum_{\gamma\in {\bs\Gamma}_{n_1,n_2}(0,0)}
\big({\rm Area}\,(\gamma)\big)^4
=\frac{(2(n_1+n_2))!}{n_1!^2 n_2!^2}\;\frac{n_1 n_2}{15}
\big({7 n_1n_2}
-({n_1+n_2})
\big)\; ,
\label{eq:area4-nxy-00}
\end{equation}
that leads to the formula (1.6) of ref.~\cite{MingoN1} after summing
over $0\le n_1\le n$ (with $n_2=n-n_1$). Interestingly, the formulas
for the moments are somewhat simpler if one considers only random
paths with fixed numbers of hops in the $+x_1$ and $+x_2$ directions,
rather that all the random paths that have a fixed length. Indeed,
besides the expected combinatorial factor, these formulas seem to
involve a polynomial in $n_{1,2}$, instead of a rational fraction in
$n=n_1+n_2$.

Similarly, we have obtained the following expressions for the moments
of order 6, 8 and 10~:
\begin{eqnarray}
&&
\smash{\sum_{\gamma\in {\bs\Gamma}_{n_1,n_2}(0,0)}}
\big({\rm Area}\,(\gamma)\big)^6
=\frac{(2(n_1+n_2))!}{n_1!^2 n_2!^2}\;\frac{n_1 n_2}{21}
\big(31 (n_1n_2)^2\nonumber\\
&&\qquad\qquad\qquad
-15n_1n_2(n_1+n_2)
+2(n_1+n_2)^2
-({n_1+n_2})
\big)\; ,
\label{eq:area6-nxy-00}
\end{eqnarray}
\begin{eqnarray}
&&
\smash{\sum_{\gamma\in {\bs\Gamma}_{n_1,n_2}(0,0)}}
\big({\rm Area}\,(\gamma)\big)^8
=\frac{(2(n_1+n_2))!}{n_1!^2 n_2!^2}\;\frac{n_1 n_2}{15}
\big(127 (n_1n_2)^3
\nonumber\\
&&\qquad\qquad\qquad
-134(n_1n_2)^2(n_1+n_2)
+53n_1n_2(n_1+n_2)^2
-6(n_1^3+n_2^3)
\nonumber\\
&&\qquad\qquad\qquad
-40n_1n_2(n_1+n_2)+8(n_1+n_2)^2-3(n_1+n_2)
\big)\; ,
\label{eq:area8-nxy-00}
\end{eqnarray}
\begin{eqnarray}
&&
\smash{\sum_{\gamma\in {\bs\Gamma}_{n_1,n_2}(0,0)}}
\big({\rm Area}\,(\gamma)\big)^{10}
=\frac{(2(n_1+n_2))!}{n_1!^2 n_2!^2}\;\frac{n_1 n_2}{33}
\big(2555 (n_1n_2)^4
\nonumber\\
&&\qquad\qquad\qquad
-4778(n_1n_2)^3(n_1+n_2)
+3745(n_1n_2)^2(n_1+n_2)^2
\nonumber\\
&&\qquad\qquad\qquad
-5290(n_1n_2)^2(n_1+n_2)-1282n_1n_2(n_1^3+n_2^3)
\nonumber\\
&&\qquad\qquad\qquad
+1918n_1n_2(n_1+n_2)^2+120(n_1^2-n_2^2)^2
\nonumber\\
&&\qquad\qquad\qquad
-1403n_1n_2(n_1+n_2)-300(n_1^3+n_2^3)
\nonumber\\
&&\qquad\qquad\qquad
+270(n_1+n_2)^2-85(n_1+n_2)
\big)\; .
\label{eq:area10-nxy-00}
\end{eqnarray}
These observations suggest that we have in general
\begin{equation}
\sum_{\gamma\in {\bs\Gamma}_{n_1,n_2}(0,0)}
\big({\rm Area}\,(\gamma)\big)^{2k}
=\frac{(2(n_1+n_2))!}{n_1!^2 n_2!^2}\;{\cal P}_{2k}(n_1,n_2)\; ,
\label{eq:poly-n12}
\end{equation}
where ${\cal P}_{2k}(n_1,n_2)$ is a polynomial of degree $2k$,
symmetric in $(n_1,n_2)$, proportional to $n_1n_2$,with rational
coefficients\footnote{Such a polynomial has $k^2$ independent
  coefficients. If we evaluate the areas of all the closed paths with
  $n_1,n_2$ such that $n_1+n_2\le n$, we get $p^2$ independent
  constraints if $n=2p$ and $p(p+1)$ independent constraints if
  $n=2p+1$. Therefore, in order to determine uniquely the polynomial
  ${\cal P}_{2k}$, it is sufficient to consider all the closed paths up
  to the length $2n=4k$. This is why one can obtain all the moments up
  to $2k=10$ by considering all the paths up to the length 20.}. The
leading term of this polynomial is of the form $c_k (n_1n_2)^{k-1}$,
and the first five terms obtained above are consistent with (defining
$c_0\equiv 1$)
\begin{equation}
\sum_{k=0}^\infty c_k\,\frac{z^{2k}}{(2k)!}=\frac{z}{\sin(z)}\; ,
\end{equation}
in agreement\footnote{In order to establish this connection, it is
  sufficient to notice that 
  \begin{equation*}
    \sum_{n_1=0}^n \frac{(2n)!}{n_1!^2(n-n_1)!^2}\;n_1^k=\frac{(2n)!^2}{n!^4}
    \;\Big[\left(\frac{n}{2}\right)^k+\mbox{subleading terms in }n\Big]\; .
  \end{equation*}} with the eq.~(1.7) of
ref.~\cite{MingoN1}.  In a separate paper, ref.~\cite{EpelbGW3}, we prove
eq.~(\ref{eq:poly-n12}) that gives the general structure of the
moments. The approach used in this paper also provides an algorithm
for calculating explicit the polynomial ${\cal P}_{2k}$ for small
values of $k$, and we have checked the formulas
(\ref{eq:area-nxy-00}), (\ref{eq:area4-nxy-00}),
(\ref{eq:area6-nxy-00}), (\ref{eq:area8-nxy-00}) and
(\ref{eq:area10-nxy-00}).

\section{Link with the almost-Mathieu operator}
\label{app:mathieu}
One can relate the statistics of closed random walks on
${\mathbbm Z}^2$ to Euclidean lattice scalar QED on a 2-dimensional
lattice by means of the discrete worldline formalism. In order to keep
track of the areas enclosed by these random walks, one should add a
magnetic field transverse to the plane in which the charged scalar
particles live. Moreover, since we would like to distinguish the
random walks according to the number of hops they make in the $x$ and
$y$ directions, we need a rectangular lattice, with distinct lattice
spacings $a_1$ and $a_2$ in the two directions. In order to write
explicit expressions, we must choose a gauge for the magnetic field. A
convenient choice is
\begin{equation}
A_1(i,j) = 0\quad,\qquad A_2(i,j)=B x = B a_1 i\; ,
\end{equation}
where $(i,j)$ are the integers that label the position of a point on
the lattice. In the lattice formulation, this background
electromagnetic field is implemented by gauge links\footnote{For
  simplicity, the electrical charge is taken to be $e=1$.}
\begin{equation}
U_1(i,j)=1\quad,\qquad U_2(i,j)=e^{iBa_2a_1i}\; .
\end{equation}
In the following, we denote $2\pi\beta\equiv Ba_1a_2$ the magnetic flux
through an elementary plaquette of the lattice.  With this gauge
choice, the action of the inverse propagator on a test function reads
\begin{equation}
(G^{-1}\phi)_{i,j}=\frac{\phi_{i+1,j}+\phi_{i-1,j}-2\phi_{i,j}}{a_1^2}
+\frac{e^{i2\pi\beta i}\phi_{i,j+1}+e^{-i2\pi\beta i}\phi_{i,j-1}-2\phi_{i,j}}{a_2^2}\; .
\label{eq:eigen-1}
\end{equation}
Moreover, the worldline representation of the propagator at equal
points (the choice of the point $0$ is arbitrary, and irrelevant
since the magnetic field is homogeneous) reads
\begin{equation}
G(0,0)
=
-\frac{{\bf a}^2}{4}
\sum_{n_1,n_2=0}^\infty\frac{h_1^{2n_1}h_2^{2n_2}}{4^{2(n_1+n_2)}}
\sum_{\gamma\in{\bs\Gamma}_{n_1,n_2}(0,0)}
\Big[\prod_{\ell\in\gamma}U_\ell\Big]\; ,
\end{equation}
where we have defined 
\begin{equation}
\frac{2}{{\bf a}^2}\equiv\frac{1}{a_1^2}+\frac{1}{a_2^2}\quad,\qquad 
h_{1,2}\equiv \frac{{\bf a}^2}{a_{1,2}^2}\; .
\end{equation}
(Note that $h_1+h_2=2$.) ${\bs\Gamma}_{n_1,n_2}(0,0)$ is the set of the
closed paths (from $(0,0)$ to $(0,0)$) drawn on the lattice, that have
exactly $n_1$ hops in the $+x$ direction and $n_2$ hops in the $+y$
direction (and therefore also $n_{1,2}$ hops in the $-x$ and $-y$
directions). The product of the link variables encountered along the
closed path is also the exponential of the magnetic flux,
\begin{equation}
\prod_{\ell\in\gamma}U_\ell
=
e^{i2\pi\beta\, {\rm Area}\,(\gamma)}\; ,
\end{equation}
where ${\rm Area}\,(\gamma)$ is the algebraic area enclosed by the
path $\gamma$. Therefore, we have
\begin{equation}
G(0,0)
=
-\frac{{\bf a}^2}{4}
\sum_{n_1,n_2=0}^\infty\frac{h_1^{2n_1}h_2^{2n_2}}{4^{2(n_1+n_2)}}
\sum_{\gamma\in{\bs\Gamma}_{n_1,n_2}(0,0)}
e^{i2\pi\beta\, {\rm Area}\,(\gamma)}
\; ,
\label{eq:G00}
\end{equation}
and one can view the diagonal elements of the propagator in a magnetic
field as a generating function for the distribution of the areas of
closed loops on the lattice.

The links $\exp(i 2\pi\beta i)$ depend only on the
$i$ coordinate. Therefore, one can perform a Fourier transform on the
$j$ coordinate. Let us define~:
\begin{equation}
\phi_{i,k}\equiv \sum_{j=0}^{N-1} \phi_{i,j}\; e^{-i2\pi \frac{kj}{N}}\; ,
\end{equation}
where $N$ is the number of lattice spacings in the $j$ direction. The
conjugate index $k$ is also an integer in the range $[0,N-1]$. The
inverse Fourier transform reads
\begin{equation}
\phi_{i,j}= \frac{1}{N}\sum_{k=0}^{N-1} \psi_{i,k}\; e^{i2\pi \frac{kj}{N}}\; ,
\end{equation}
If we consider an infinitely large lattice, $N\to+\infty$, one can use
a continuous momentum variable $\nu\equiv 2\pi k/N$, so that the above
equation becomes
\begin{equation}
\phi_{i,j}=\int_0^{2\pi}\frac{d\nu}{2\pi}\;\psi_{i,\nu}\;e^{i\nu j}\; .
\end{equation}
By inserting this equation into eq.~(\ref{eq:eigen-1}), we obtain
\begin{equation}
\Big[\big({\bf a}^2 G^{-1}+4\big)\psi\Big]_{i,\nu}
=\sum_{i'}\int\frac{d\nu'}{2\pi}\;H^{(\beta)}_{i\nu,i'\nu'}\;\psi_{i',\nu'}\; ,
\label{eq:eigen-2}
\end{equation}
where $H^{(\beta)}$ is known as the (anisotropic) almost-Mathieu
operator~:
\begin{equation}
\left(H^{(\beta)}\right)_{i\nu,i'\nu'}
\equiv
\Big[h_1(\delta_{i,i'+1}+\delta_{i,i'-1})
+
2\,h_2\cos(2\pi\beta i+\nu)\,\delta_{i,i'}\Big]\;2\pi\delta(\nu-\nu')\; .
\label{eq:H-1}
\end{equation}
This Hamiltonian has been the subject of many studies, both for its
practical interest in models of the quantum Hall
effect~\cite{Harpe1,Harpe2,WanniO1,ChmelO1}, and for its intrinsic mathematical
interest~\cite{Hofst1,LamouM1,Lamou1,LamouMP1,AvilaD1,Avila1,Daman1,Jitom1}
as an example of quasi-periodic Hamiltonian (when $\beta$ is an
irrational number).

From eq.~(\ref{eq:eigen-2}), we get that
\begin{equation}
G={\bf a}^2 \big[H^{(\beta)}-4\big]^{-1}=-\frac{{\bf a}^2}{4}\sum_{n=0}^\infty
\left(\frac{H^{(\beta)}}{4}\right)^n\; .
\label{eq:prop-Hn}
\end{equation}
Since the external magnetic field is homogeneous, the diagonal
elements of the propagator can be obtained by taking the trace
(divided by the number of lattice sites in the $i$ direction). Thus,
by comparing eqs.~(\ref{eq:G00}) and (\ref{eq:prop-Hn}) we recover the
well-known connection\cite{BeguiVZ1,BelliCBC1,MashkO1} between the
trace of $\big(H^{(\beta)}\big)^n$ and the distribution of the areas
of the closed random walks of length $n$,
\begin{equation}
\sum_{n_1+n_2=n}{h_1^{2n_1}h_2^{2n_2}}
\!\!\!\!\sum_{\gamma\in{\bs\Gamma}_{n_1,n_2}(0,0)}
\!\!\!\!e^{i2\pi\beta\, {\rm Area}\,(\gamma)}
=
\lim_{N\to\infty}\frac{1}{N}\sum_{i=1}^{N}\int_0^{2\pi}\frac{d\nu}{2\pi}\;
\Big[\big(H^{(\beta)}\big)^{2n}\Big]_{i\nu,i\nu}\, .
\end{equation}

\section{$\big<\phi_a^*(0)\phi_a(0)\big>$ from lattice perturbation theory}
\label{app:LPT}
In this appendix, we calculate $\big<\phi^*_a(x) \phi_a(x)\big>$ in
lattice perturbation theory in order to make the connection with our
results obtained in the lattice worldline representation. In lattice
perturbation theory (see ref.~\cite{Capit1} for a comprehensive
review), the expectation value of any operator ${\cal O}$ can be
calculated perturbatively according to
\begin{equation}
\big< {\cal O}\big> = \frac{\big[\int D\phi D\phi^\dagger\big]\;
 {\cal O}[\phi,\phi^\dagger]\; e^{-S[\phi^\dagger,\phi]}}
{\int \big[D\phi D\phi^\dagger\big]\; e^{-S[\phi^\dagger,\phi]}}\; .
\end{equation}
In our case, the action reads
\begin{equation}
S[\phi^\dagger,\phi]\equiv S_0[\phi^\dagger,\phi]+S_{\rm int}[\phi^\dagger,\phi,U]\;,
\end{equation}
with
\begin{eqnarray}
S_0[\phi^\dagger,\phi]&=&
-V_a\sum\limits_x \phi_x^\dagger \sum\limits_{r=1}^d 
\frac{1}{a_r^2}
\left[\phi({x+a_r \hat r})+\phi({x-a_r \hat r}) -2 \phi(x)\right]\;,\\
S_{\rm int}[\phi^\dagger,\phi]
&=&
-V_a\sum\limits_x \sum\limits_{r=1}^d \frac{1}{a_r^2}
\Big[\phi^\dagger(x) (U_{r}(x)-\mathbb{I})\phi({x+a_r \hat r})
\nonumber\\
&&\qquad\qquad\qquad
+
\phi^\dagger({x+a_r\hat r}) (U_{r}^{-1}(x)-\mathbb{I})\phi({x})\Big]\;,
\end{eqnarray}
and $V_a \equiv \prod\limits_{r=0}^d a_i$. In the limit of an
infinitely large lattice, the free propagator reads
\begin{eqnarray}
G_0(x,y)
 &&= \left[\prod\limits_{r=1}^d\int_{-\frac{\pi}{a_r}}^{\frac{\pi}{a_r}} \frac{dp_r}{2\pi}e^{i p_r (x_r-y_r)}\right]\frac{1}{2\sum\limits_{l=1}^d\frac{1}{a_l^2}[1-\cos (p_l a_l)]}\nonumber\\
&&=\frac{1}{V_a}\left[\prod\limits_{r=1}^d\int_{-\pi}^{\pi} \frac{dp_r}{2\pi}e^{i p_r (m_r-n_r)}\right]\underbrace{\frac{1}{2\sum\limits_{l=1}^d\frac{1}{a_l^2}(1-\cos p_l )}}_{\widehat{G}(\{p_l\})}\; ,
\end{eqnarray}
with $x_r=m_r a_r, y_r = n_r a_r$. Since it is a Green's function of
the discrete Euclidean d'Alembertian operator, it satisfies the
identity
\begin{equation}
-\sum\limits_{r=1}^d \frac{1}{a_r^2}\left[G_0({x+a_r \hat r},y)+G_0({x-a_r \hat r},y) -2 G_0(x,y)\right]=\frac{1}{V_a}\delta_{
m_r,n_r}\; .
\end{equation}

Let us now calculate the first two terms of the expansion of
$\big<\phi^*_a(0)\phi_a(0)\big>$ in powers of the external field,
\setbox4\hbox to
1.5cm{\resizebox*{1.5cm}{!}{\includegraphics{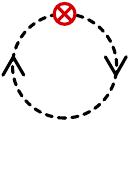}}} \setbox5\hbox to
1.5cm{\resizebox*{1.5cm}{!}{\includegraphics{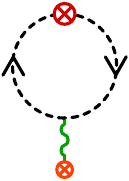}}} \setbox6\hbox to
1.5cm{\resizebox*{1.5cm}{!}{\includegraphics{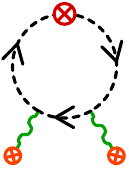}}}
\begin{equation}
\left<\phi^*_a(0)\phi_a(0)\right>=
\raise -12mm\box4
+
\raise -12mm\box5
+
\raise -12mm\box6
+\cdots\equiv I_0+I_1+I_2+\cdots\;.
\end{equation}
The term of order zero is given by~:
\setbox5\hbox to 1.5cm{\resizebox*{1.5cm}{!}{\includegraphics{I0}}}
\begin{eqnarray}
I_0
&\equiv&
\raise -12mm\box5
=\null\frac{{\rm tr}_{{\rm adj}}(1)}{2dV_a}\int_{-\pi}^{\pi} \frac{d^d p}{(2\pi)^d}
\frac{1}{{\bf a}^{-2}-\frac{1}{d}\sum\limits_{l=1}^d a_l^{-2}\cos p_l}
\nonumber\\
&=& \frac{{\bf a}^{2}~\text{tr}_{\text{adj}}(1)}{2dV_a}\int_0^\infty dt \;e^{-t} 
\int_{-\pi}^{\pi} \frac{d^d p}{(2\pi)^d}\;
\exp\left({\sum\limits_{l=1}^d\frac{h_r t}{d}\cos p_l}\right)
\nonumber\\
&=&\frac{{\bf a}^{2}~\text{tr}_{\text{adj}}(1)}{2dV_a}\int_0^\infty dt\; e^{-t}\;
\prod\limits_{r=1}^d
I_0(\tfrac{h_r t}{d})\; ,
\end{eqnarray}
which is equivalent to eqs.~(\ref{eq:exp-phi2-3}) and
(\ref{eq:C0-bessel-aniso}).

Next, we calculate the corrections to
$\big<\phi^*_a(0)\phi_a(0)\big>$ of order ${\cal O}(g^2 a^2)$. By using the power
series expansion of the link variable,
\begin{equation}
U_{r}(x)=e^{-i g a_r A_r(x)}=\mathbb{I}-ig a_r A_r(x)-\frac{g^2}{2}a_r^2 A^2_r(x)+\cdots\; ,
\end{equation}
and keeping only terms of ${\cal O}(g^2)$, we find 
\begin{equation}
I_1+I_2
=\frac{g^2}{2}\int\frac{d^dk_1}{(2\pi)^d}\frac{d^dk_2}{(2\pi)^d}
\;\sum\limits_{r,s=1}^d \widetilde{A}_r^a(k_1)\widetilde{A}_s^a(k_2)I_{rs}(k_1,k_2) 
+ {\cal O}(g^3)\;,
\end{equation}
with
\begin{eqnarray}
&&I_{rs}(k_1,k_2)=
\frac{1}{V_a}\int_{-\pi}^\pi \frac{d^dp}{(2\pi)^d}\;
\widehat{G}(\{p_l\})
\Big[\frac{1}{a_ra_s}
\widehat{G}(\{p_l+k_{1l} a_l\})
\widehat{G}(\{p_l-k_{2l} a_l\}) 
\nonumber\\
&&
\!\times\!
\left(
e^{i (p_r-p_s)}
\!-\!
e^{i p_r+i p_s-i k_{2s}a_s}
\!+\!
e^{-i (p_r+k_{1r} a_r)+i (p_s-k_{2s} a_s)}
\!-\!
e^{-i (p_r+k_{1r} a_r)-i p_s}\right)\nonumber\\
&&
\qquad\qquad
-\frac{\delta_{rs}}{2}
\widehat{G}(\{p_l+(k_{1l}+k_{2l}) a_l\})
(e^{i p_r}+e^{-i p_r-i(k_{1r}+k_{2r}) a_r})\Big]\;.
\end{eqnarray}
The terms of order $g^2 a^2$ can be obtained by performing a series
expansion of $I_{rs}$ in terms of $k_{1,2} a$. It is easy to show that
\begin{eqnarray}
I_{rs}(k_1,k_2)=I_{1rs}+I_{2,1rs}+I_{2,2rs}+I_{2,3rs}+I_{2,4rs}+{\cal O}(a^3)\;,
\end{eqnarray}
with
\begin{eqnarray}
&&
I_{1rs}= -\frac{\delta_{rs}}{2V_a}
\sum_{i=1}^d\int_{-\pi}^\pi \frac{d^dp}{(2\pi)^d}\;
(k_{1i}+k_{2i})^2 a_i^2\;\widehat{G}(\{p_l\})
\frac{\partial^2\widehat{G}(\{p_l\})}{\partial p_i^2}\;\cos p_r\;,
\nonumber
\end{eqnarray}
\begin{eqnarray}
&&
I_{2,1rs}=\frac{1}{V_a}\sum_{i,j=1}^d\int_{-\pi}^\pi 
\frac{d^dp}{(2\pi)^d}\;
\frac{2 a_i a_j}{a_ra_s}\;
(k_{1i}k_{1j}+k_{2i}k_{2j})\;
 \widehat{G}^2(\{p_l\})\nonumber\\
&&\qquad\qquad\qquad\qquad\qquad\qquad
\times\frac{\partial^2\widehat{G}(\{p_l\})}{\partial p_i \partial p_j} \sin p_r \sin p_s\;,
\nonumber
\end{eqnarray}
\begin{eqnarray}
&&
I_{2,2rs}=-\frac{4}{V_a}\sum_{i,j=1}^d
\int_{-\pi}^\pi \frac{d^dp}{(2\pi)^d}\;\frac{a_i a_j}{a_ra_s}\;
k_{1i} k_{2j} \widehat{G}(\{p_l\})
\nonumber\\
&&\qquad\qquad\qquad\qquad\qquad\qquad
\times\frac{\partial\widehat{G}(\{p_l\})}{\partial p_i}  
\frac{\partial\widehat{G}(\{p_l\})}{\partial p_j}\;  \sin p_r \sin p_s\;,\nonumber
\end{eqnarray}
\begin{eqnarray}
&&
I_{2,3rs}=\frac{2}{V_a}\int_{-\pi}^\pi \frac{d^dp}{(2\pi)^d}\;
\widehat{G}^2(\{p_l\})\;
\Big[(k_{1s}-k_{2s}) k_{1r}\;
\frac{\partial\widehat{G}(\{p_l\})}{\partial p_s} \sin p_s \cos p_r\nonumber\\
&&\qquad\qquad\qquad\qquad\qquad\qquad
-(k_{1r}-k_{2r}) k_{2s}\;
\frac{\partial\widehat{G}(\{p_l\}) }{\partial p_r}\cos p_s \sin p_r\Big]\;,
\nonumber
\end{eqnarray}
\begin{eqnarray}
&&I_{2,4rs}=-\frac{1}{V_a}\int_{-\pi}^\pi \frac{d^dp}{(2\pi)^d}\;
\widehat{G}^3(\{p_l\})\;
\Big[\delta_{rs}(k_{1r}^2+k_{2r}^2) \sin^2 p_r\nonumber\\
&&\qquad\qquad\qquad\qquad\qquad\qquad
+ k_{1r} k_{2s}\left(\cos p_r \cos p_s+\sin^2 p_r \delta_{sr}\right)\Big]\;.
\end{eqnarray}
In order to simplify the above integrals, we need the following
identities
\begin{eqnarray}\label{eq:Gidentity}
\frac{\partial\widehat{G}(\{p_l\})}{\partial p_r} 
&=& -\frac{2}{a_r^2} \sin p_r\; \widehat{G}^2(\{p_l\})\;,\nonumber\\
\frac{\partial^2 \widehat{G}(\{p_l\})}{\partial p_r \partial p_s} 
&=& \frac{8}{a_r^2 a_s^2} \sin p_r \sin p_s\;\widehat{G}^3(\{p_l\})
-\frac{2}{a_r^2} \cos p_r\; \widehat{G}^2(\{p_l\}) \delta_{rs}\;,\nonumber\\
&&
=2\widehat{G}^{-1}(\{p_l\})
\frac{\partial \widehat{G}(\{p_l\})}{\partial p_r} 
\frac{\partial\widehat{G}(\{p_l\})}{\partial p_s}
-\frac{2}{a_r^2} \cos p_r \;\widehat{G}^2(\{p_l\}) \delta_{rs}\;.\nonumber\\
&&
\end{eqnarray}
Moreover, the terms that are total derivatives with respect to any
component of $p$ in the above integrands always give vanishing
contributions. After some algebra we find
\begin{eqnarray}
&&
I_{1rs}=\frac{\delta_{rs}}{3V_a}\int_{-\pi}^\pi \frac{d^dp}{(2\pi)^d}\;
\widehat{G}^3(\{p_l\})\;
\Big[\sum_{i=1}^d(k_{1i}+k_{2i})^2 \cos p_i \cos p_r \nonumber\\
&&\qquad\qquad\qquad\qquad\qquad\qquad
+ 2 (k_{1r}+k_{2r})^2 \sin^2 p_r\Big]\;,\nonumber
\end{eqnarray}
\begin{eqnarray}
&&
I_{2,1rs}=\frac{1}{3V_a}\int_{-\pi}^\pi \frac{d^dp}{(2\pi)^d}\;
\widehat{G}^3(\{p_l\})\;
\Big[2(k_{1r} k_{1s}+k_{2r}k_{2s})\cos p_r\cos p_s \nonumber\\
&& \qquad\qquad
- (k_{1r}^2+k_{2r}^2)\delta_{rs}\sin^2 p_r  
- \sum\limits_{i=1}^d(k_{1i}^2+k_{2i}^2)\cos p_i \cos p_r \delta_{rs}\Big]\;,\nonumber
\end{eqnarray}
\begin{eqnarray}
&&
I_{2,2rs}=-\frac{1}{3V_a}\int_{-\pi}^\pi \frac{d^dp}{(2\pi)^d}\;
\widehat{G}^3(\{p_l\})\;
\Big[(k_{1r}k_{2s} +k_{1s}k_{2r}) \cos p_r\cos p_s \nonumber\\
&& \qquad\qquad
- 3k_{1r}k_{2r}\delta_{rs} \sin^2 p_r  
+ \sum\limits_{i=1}^d k_{1i} k_{2i}\cos p_i \cos p_r \delta_{rs}\Big]\;,\nonumber
\end{eqnarray}
\begin{eqnarray}
&&I_{2,3rs}=-\frac{2}{3V_a}\!\int_{-\pi}^\pi \!\frac{d^dp}{(2\pi)^d}\,
\widehat{G}^3(\{p_l\})\,
\Big[(k_{1s}k_{1r}\!+\!k_{1s}k_{1r}\!-\!2k_{1r} k_{2s} )\cos p_s \cos p_r \nonumber\\
&&\qquad\qquad
-(k_{1r}^2+k_{2r}^2-2k_{1r} k_{2r} )\delta_{rs}\sin^2 p_r\Big]\;.
\end{eqnarray}
As a result, up to ${\cal O}(g^2 a^2)$, we have
\begin{eqnarray}
I_{rs}(k_1,k_2)&=&
\frac{1}{3V_a}\int_{-\pi}^\pi \frac{d^dp}{(2\pi)^d}\;
\widehat{G}^3 (\{p_l\})\;
\Big[\delta_{rs} \sum\limits_{i=1}^dk_{1i} k_{2i}\cos p_i \cos p_r  
\nonumber\\
&&\qquad\qquad\qquad\qquad\qquad\quad
- k_{1s} k_{2r}\cos p_r \cos p_s \Big]\;,
\end{eqnarray}
and
\begin{eqnarray}
I_1+I_2&=&
-\frac{g^2}{12V_a}\!\sum_{r,s} [ \partial_r A^a_s(x) - \partial_s A^a_r(x)]^2
\!\int_{-\pi}^\pi\! \frac{d^dp}{(2\pi)^d}\,
\widehat{G}^3 (\{p_l\})\, \cos p_r \cos p_s\nonumber\\
&&=-\frac{g^2 {\bf a}^6}{192 d^3 V_a}\sum_{r\neq s} [ \partial_r A^a_s(x) - \partial_s A^a_r(x)]^2
\nonumber\\
&&\qquad\qquad\qquad\times
\int_0^\infty dt\;t^2\; e^{-t}\;
I_1(\tfrac{h_r t}{d})\;
I_1(\tfrac{h_s t}{d})\;
\prod\limits_{i\neq r,s}I_0(\tfrac{h_i t}{d})\;,\nonumber
\end{eqnarray}
where we have used the integral
\begin{eqnarray}
\frac{1}{x^3}=\frac{1}{2}\int_0^\infty dt\;t^2\; e^{-t x}\;.
\end{eqnarray}
In summary, we have 
\begin{eqnarray}
\big< \phi_a^*(0)\phi_a(0)\big> &=&
\frac{{\bf a}^{2}\,{\rm tr}_{\rm adj}(1)}{2dV_a}\int_0^\infty dt\; e^{-t}\;
\prod_{r=1}^d
I_0(\tfrac{h_r t}{d})
\nonumber\\
&&
-\frac{g^2 {\bf a}^6}{192 d^3 V_a}\sum_{r\neq s} [ \partial_r A^a_s(x) - \partial_s A^a_r(x)]^2
\nonumber\\
&&\qquad\!\!\!\times
\int_0^\infty \!\!dt\;t^2\; e^{-t}\;
 I_1(\tfrac{h_r t}{d})\,I_1(\tfrac{h_s t}{d})\,
\prod_{i\neq r,s}I_0(\tfrac{h_i t}{d})+{\cal O}(g^2 a^3)\; .
\nonumber\\
&&
\end{eqnarray}
This result is identical to the formulas obtained in the worldline
formalism for this expectation value (see the
eqs.~(\ref{eq:exp-phi2-3}) and (\ref{eq:C4-bessel-aniso})). Therefore,
one can view this alternate derivation as an indirect analytical proof
of the formula (\ref{eq:area-nxy-00}), that we have used in obtaining
the next-to-leading order coefficient in the anisotropic case. In
principle, one could apply the same technique in order to prove all
the other formulas conjectured in the appendix \ref{app:area-aniso},
although this appears to be a rather intricate task.

\section{Continuous time and discrete space}
\label{app:cont-time}
The extreme case of anisotropy is realized when one of the coordinates
(time in our discussion) is treated as a continuous variable, while
the others remain discretized on a lattice of spacing $a$. In order to
simplify the treatment of the background field, it is useful to
assume that the temporal gauge is used, $A_0=0$. Therefore, the
inverse propagator is
\begin{equation}
D^2\equiv g_{_{00}}\partial_0^2 - \sum_{i=1}^{d_s}\nabla^+_i\nabla^-_i\; ,
\end{equation}
where $\nabla_i^\pm$ are the forward and backward discrete covariant
derivatives on a grid of lattice spacing $a$, and $d_s\equiv d-1$ the number
of spatial dimensions. At this point, we have also kept undetermined
the $00$ component of the metric tensor, $g_{_{00}}$, in order to discuss
later the difference in this formalism between the Minkowski and the
Euclidean metric.

For the purpose of this discussion, we can first ignore the background
field completely, and reintroduce it later via the Wilson loop made of
the gauge links accumulated along the random walk. Let us recall that
\begin{equation}
\nabla^+_i\nabla^-_i f(x)=\frac{f(x+\hat\imath)+f(x-\hat\imath)}{a^2}-\frac{2\,f(x)}{a^2}\; .
\end{equation}
It is convenient to write the inverse propagator as follows
\begin{equation}
D^2 = \frac{2d_s}{a^2}(\tfrac{g_{_{00}}a^2}{2d_s} \partial_0^2+1)
\Big[1-\left(\tfrac{g_{_{00}}a^2}{2d_s} \partial_0^2+1\right)^{-1}{\bs H}\Big]\; ,
\label{eq:invD2}
\end{equation}
where we denote
\begin{equation}
{\bs H}f(x)\equiv\sum_{i=1}^{d_s}\frac{f(x+\hat\imath)+f(x-\hat\imath)}{2d_s}
\end{equation}
the operator ${\bs H}$ generates the hops for a random walk on a
square lattice in $d_s$ dimensions. Using eq.~(\ref{eq:invD2}), we can
write the inverse propagator as
\begin{eqnarray}
\frac{2d_s}{a^2}\,\frac{1}{D^2}
&=&
(\tfrac{g_{_{00}}a^2}{2d_s} \partial_0^2+1)^{-1}
+
(\tfrac{g_{_{00}}a^2}{2d_s} \partial_0^2+1)^{-1}{\bs H}(\tfrac{g_{_{00}}a^2}{2d_s} \partial_0^2+1)^{-1}
\nonumber\\
&&
+(\tfrac{g_{_{00}}a^2}{2d_s} \partial_0^2+1)^{-1}{\bs H}(\tfrac{g_{_{00}}a^2}{2d_s} \partial_0^2+1)^{-1}{\bs H}(\tfrac{g_{_{00}}a^2}{2d_s} \partial_0^2+1)^{-1}
+\cdots
\nonumber\\
&&
\label{eq:diff-xt}
\end{eqnarray}
In this form, the inverse propagator appears as a sum of terms, each
of which is an alternating product of ${\bs H}$ (i.e. single hops on
the lattice that represents space) and of the inverse of the operator
$\tfrac{g_{_{00}}a^2}{2d_s} \partial_0^2+1$. We shall now rewrite this
object in a way that clarifies its physical meaning. Let us start
with its heat kernel representation
\begin{equation}
\frac{1}{\tfrac{g_{_{00}}a^2}{2d_s} \partial_0^2+1}
=
\int_0^\infty dt\;
\exp\Big(-t\left(1+\tfrac{g_{_{00}}a^2}{2d_s} \partial_0^2\right)\Big)\; .
\label{eq:invD2-t}
\end{equation}
The second step is the following identity
\begin{equation}
e^{\tfrac{\alpha}{2}\partial^2}
=
\int_{-\infty}^{+\infty} \frac{dz}{\sqrt{2\pi\alpha}}\;e^{-\tfrac{z^2}{2\alpha}}\; e^{z\partial}\; ,
\end{equation}
or more explicitly
\begin{equation}
e^{\tfrac{\alpha}{2}\partial^2}
f(x)
=
\int_{-\infty}^{+\infty} \frac{dz}{\sqrt{2\pi\alpha}}\;e^{-\tfrac{z^2}{2\alpha}}\; f(x+z)\; .
\end{equation}
In words, the operator $\exp(\tfrac{\alpha}{2}\partial^2)$ is a
diffusion operator that smears the target function by convolution with
a Gaussian of variance $\alpha$. But this interpretation is only
possible if $\alpha$ is a positive real number. The closest formula if
$\alpha$ is negative would be
\begin{equation}
e^{\tfrac{\alpha}{2}\partial^2}
f(x)
\empile{=}\over{\alpha<0}
\int_{-\infty}^{+\infty} \frac{dz}{\sqrt{2\pi|\alpha|}}\;e^{-\tfrac{z^2}{2|\alpha|}}\; f(x+iz)\; ,
\end{equation}
but it requires that we complexify the variable $x$. This is precisely
the situation we face when employing this formula to rewrite the
inverse propagator: $g_{_{00}}$ must be negative in order to obtain a
properly normalized Gaussian. If we start from the Minkowski metric
($g_{_{00}}=+1$), we can still get a Gaussian with the ``correct
sign'' if we complexify the time. The conclusion of this digression
is that the diffusive interpretation of the time evolution of a
quantum system is only possible with imaginary time. 

From now on, we assume that $g_{_{00}}=-1$. Eq.~(\ref{eq:invD2-t}) can
be rewritten as
\begin{equation}
\frac{1}{1-\tfrac{a^2}{2d_s} \partial_0^2}
f(x_0)
=
\int_0^\infty dt\;\sqrt{\frac{d_s}{2\pi t a^2}}\int_{-\infty}^{+\infty}dz\;
e^{-t}\;
e^{-\frac{d_s z^2}{2 a^2 t}}\;
f(x_0+z)\; .
\label{eq:invD2-t-1}
\end{equation}
In the right hand side, we recognize an integral over diffusion
processes of ``duration'' $t$, weighted by a factor $\exp(-t/2)$ that
suppresses the contribution of diffusions longer than $t\sim 1$. Under
such a diffusion, the time $x_0$ can shift by an amount of order $a$
(the spatial lattice spacing).

The interpretation of eq.~(\ref{eq:diff-xt}) is now quite
transparent. The inverse of $D^2$ is obtained by intertwining
continuous diffusions in imaginary time (of arbitrary lengths, but
weighted in such a way that the displacement in time is of order $a$)
and single hops in one of the spatial directions. Note that in
eq.~(\ref{eq:invD2-t-1}), the integration over the length $t$ can be
done analytically by using
\begin{equation}
\int_0^\infty \frac{dt}{\sqrt{t}}\;e^{-t}\;e^{-c/t}=\sqrt{\pi}\;e^{-2\sqrt{c}}\; .
\end{equation}
Therefore, we can write
\begin{equation}
\frac{1}{1-\tfrac{a^2}{2d_s} \partial_0^2}
f(x_0)
=
\frac{\gamma}{2}\int_{-\infty}^{+\infty}dz\;e^{-\gamma|z|} \;f(x_0+z)\; ,
\end{equation}
where we denote $\gamma\equiv\sqrt{2d_s/a^2}$. We see that the
application of this operator is a convolution with a Laplace
distribution. The typical width of the resulting smearing is
$z\sim\gamma^{-1}\sim a$. In other words, the {\sl spatial} lattice
spacing also controls the typical size of the jumps in time that are
interspersed between the jumps on the spatial lattice. This is
consistent with our conclusion of the section \ref{sec:aniso}, that
the results become independent of the smallest lattice spacing when it
becomes infinitesimally small (the continuous-time description adopted
in this appendix corresponds to a discrete description of time with a
time interval that goes to zero).  In particular, the ultraviolet
behavior in this limit is controlled by the next-to-smallest lattice
spacing.

In the series expansion of the inverse propagator, we need to evaluate
the $n+1$-th power of this operator\footnote{The choice of the $A_0=0$
  gauge is crucial here. Indeed, if we had a non-zero position
  dependent $A_0$ background field, the hops in the time direction
  would not commute with the spatial hops, and it would be impossible
  to collect them as the $n+1$-th power of a single temporal hop
  operator.}, acting on a starting distribution of the form
$\delta(x_0)$,
\begin{eqnarray}
\alpha_n(x_0)&\equiv& 
\left[\frac{1}{1-\tfrac{a^2}{2d_s} \partial_0^2}\right]^{n+1}\;\delta(x_0)
\nonumber\\
&=&
\left(\frac{\gamma}{2}\right)^{n+1}\!\!\!\int\!\! dz_1\cdots dz_{n+1}\;
e^{-\gamma(|z_1|+\cdots+|z_{n+1}|)}\;\delta(x_0+z_1+\cdots+z_{n+1})
\nonumber\\
&=&
\left(\frac{\gamma}{2}\right)^{n+1}\int_{-\infty}^{+\infty}\frac{du}{2\pi}\;e^{iux_0}
\left[\int dz\;e^{-\gamma|z|+iuz}\right]^{n+1}
\nonumber\\
&=&
\left(\frac{\gamma}{2}\right)^{n+1}\int_{-\infty}^{+\infty}\frac{du}{2\pi}\;e^{iux_0}
\left[\frac{2\gamma}{u^2+\gamma^2}\right]^{n+1}\; .
\end{eqnarray}
For $x_0=0$, the evaluation of the integral is straightforward and one obtains
\begin{equation}
\alpha_n(0) = \frac{\gamma}{2}\;\frac{(2n)!}{4^n\,n!^2}\; .
\label{eq:timestep}
\end{equation}
The second factor is the probability that a random walk in one
dimension returns to the origin after $2n$ steps.

Using these results, we obtain the following expression for the
leading term of $\big<\phi_a^*(0)\phi_a(0)\big>$,
\begin{equation}
\big<\phi^*(0)\phi(0)\big>
=
\frac{1}{2d_s a}\;{\widetilde{\bs C}}_0\;{\rm tr}_{\rm adj}\,(1)+\cdots\; ,
\end{equation}
with 
\begin{equation}
{\widetilde{\bs C}}_0
\equiv
\frac{\gamma}{2}\sum_{n=0}^\infty\frac{(4n)!}{4^{2n}(2n)!}\;
\frac{1}{(2d_s)^{2n}}\!\!\!\sum_{n_1+\cdots+n_{d_s}=n}\frac{(2n)!}{n_1!^2\cdots n_{d_s}!^2}\; .
\end{equation}
In this summation, we must have an even number $2n$ of spatial hops
since their combination must form a closed loop on the spatial
lattice. The first factor is therefore eq.~(\ref{eq:timestep})
evaluated for $2n$ instead of $n$. The second factor counts the number
of closed random walks of length $2n$ in $d_s$ dimensions. In order to
disentangle the sums over $n_1,\cdots,n_{d_s}$, we need a modified
form of the Borel transformation, tuned to cancel the $n$-dependent
prefactor. Notice first that
\begin{equation}
\frac{1}{\sqrt{\pi}}\int_0^\infty\frac{d\tau}{\sqrt{\tau}}\;e^{-\tau}\;
\tau^{2n}=\frac{(4n)!}{4^{2n}(2n)!}\; . 
\end{equation}
This leads immediately to
\begin{equation}
{\widetilde{\bs C}}_0
=
\frac{\gamma}{2}\frac{1}{\sqrt{\pi}}\int_0^\infty\frac{d\tau}{\sqrt{\tau}}\;e^{-\tau}\;
I_0^{d_s}(\tfrac{\tau}{d_s})\; ,
\end{equation}
and finally to
\begin{equation}
\big<\phi^*(0)\phi(0)\big>
\empile{=}\over{a_d\ll a_{1\cdots d-1}\to 0}\frac{{\rm tr}_{\rm adj}\,(1)}{2a^{d-2}}
\frac{1}{\sqrt{2\pi d_s}}\int_0^\infty\frac{d\tau}{\sqrt{\tau}}\;e^{-\tau}\;I_0^{d_s}(\tfrac{\tau}{d_s})\; ,
\end{equation}
which is identical to the formula of the footnote \ref{foot:C0}, that
was obtained by taking the limit $a_d\to 0$ in a discrete time
formulation.


\end{document}